\documentclass[a4paper,fleqn]{cas-dc}

\usepackage[numbers,square]{natbib}

\usepackage{graphicx}
\usepackage{rotating}
\usepackage{graphicx}
\usepackage{longtable}
\usepackage{multirow}

\usepackage{xurl} % allows line breaks at any character in URLs
\usepackage{hyperref}
\hypersetup{breaklinks=true}

\usepackage{lineno,hyperref}
\usepackage{algpseudocode}
\usepackage[linesnumbered, ruled]{algorithm2e}
\usepackage{amsmath}

\usepackage{array, multirow, graphicx}
\newcolumntype{C}[1]{>{\centering\arraybackslash}p{#1}}

\usepackage{amssymb}
\usepackage{caption}

\usepackage{adjustbox}
\usepackage{float}
\usepackage{makecell}
\usepackage{adjustbox}
\usepackage{multirow}
\usepackage{lineno}
\modulolinenumbers[1]

\def\tsc#1{\csdef{#1}{\textsc{\lowercase{#1}}\xspace}}
\tsc{WGM}
\tsc{QE}
\tsc{EP}
\tsc{PMS}
\tsc{BEC}
\tsc{DE}
\begin{document}
% \linenumbers
% \setlength\linenumbersep{6pt} % default is small; increase to avoid overlap
% \renewcommand\linenumberfont{\normalfont\tiny\sffamily\color{gray}}

\let\WriteBookmarks\relax
\def\floatpagepagefraction{1}
\def\textpagefraction{.001}

\shorttitle{A Self-Attention-Driven Deep Denoiser Model for Real Time Lung Sound Denoising in Noisy Environments}

\shortauthors{Shuvo et~al.}

\title [mode = title]{A Self-Attention-Driven Deep Denoiser Model for Real Time Lung Sound Denoising in Noisy Environments}

\author[1]{Samiul Based Shuvo}[orcid=0000-0002-5035-2114,]



\ead{sbshuvo@bme.buet.ac.bd}

\affiliation[1]{
    organization={m-Health Lab, Department of Biomedical Engineering, Bangladesh University of Engineering and Technology},
    addressline={Dhaka-1205, Bangladesh}
}

\author[2]{Syed Samiul Alam }[
   orcid=0000-0002-3742-6692,
   ]
\ead{syed.samiul.alam@ieee.org}
\affiliation[2]{
    organization={Department of Electronical and Electronic Engineering, Khulna University of Engineering \& Technology (KUET)},
    addressline={Khulna-9203, Bangladesh}
}

\author[1,3]{Taufiq~Hasan}[%
   orcid=0000-0002-6142-3344,
   ]
\cormark[1]
% \ead{taufiq.hasan@jhu.edu;}
\ead[URL]{taufiq.hasan@jhu.edu; taufiq@bme.buet.ac.bd}

\affiliation[3]{
    organization={Center for Bioengineering Innovation and Design, Department of Biomedical Engineering, Johns Hopkins University},
    city={Baltimore},
    state={MD}
}

\cortext[cor1]{Corresponding author}

\begin{abstract}
~\textbf{Objective:} Lung auscultation is a valuable tool in diagnosing and monitoring various respiratory diseases. However, lung sounds (LS) are significantly affected by numerous sources of contamination, especially when recorded in real-world clinical settings. Conventional denoising models prove impractical for LS denoising, primarily owing to spectral overlap complexities arising from diverse noise sources. To address this issue, we propose a specialized deep-denoiser model (Uformer) for lung sound denoising. \textbf{Methods:} The proposed Uformer model is constituted of three modules: a Convolutional Neural Network (CNN) encoder module, dedicated to extracting latent features; a Transformer encoder module, employed to further enhance the encoding of unique LS features and effectively capture intricate long-range dependencies; and a CNN decoder module, employed to generate the denoised signals. An ablation study was performed in order to find the most optimal architecture. ~\textbf{Results:} The performance of the proposed Uformer model was evaluated on lung sounds induced with different types of synthetic and real-world noises. Lung sound signals of -12 dB to 15 dB signal-to-noise ratio (SNR) were considered in testing experiments. The proposed model showed an average SNR improvement of 16.51 dB when evaluated with -12 dB LS signals. Our end-to-end model, with an average SNR of 19.31 dB, outperforms the existing model when evaluated with ambient noise and fewer parameters.~\textbf{Conclusion:} Based on the qualitative and quantitative findings in this study, it can be stated that Uformer is robust and generalized to be used in assisting the monitoring of respiratory conditions.
\end{abstract}


\maketitle

\section{Introduction}

The respiratory system plays a pivotal role in human health, making it crucial to diagnose respiratory diseases accurately and monitor patient conditions. Pulmonary disorders are the third leading cause of death in the world\cite{ihme2023}. In this context, lung sounds serve as a valuable tool for physicians. They offer insights into both the normal physiological processes and potential pathologies within the respiratory system\cite{reichert2008analysis}. Generally, these sounds can be classified into normal and abnormal categories \cite{lozano2016performance}. Auscultation, the practice of listening to physiological sounds such as those of the lungs, is a technique often used by physicians and clinicians to detect respiratory illnesses such as Asthma, Chronic Obstructive Pulmonary Disease (COPD), Bronchiolitis, and Bronchitis. Despite its significance, this method often encounters challenges due to inter-listener variability, where different individuals might interpret the sounds differently. There has been an emerging research trend toward computer-based lung sound (LS) analysis to address this subjective inconsistency. For instance, by utilizing electronic stethoscopes, this technological advancement provides a more objective and consistent method to evaluate lung functions~\cite{bohadana2014fundamentals,gurung2011computerized}.

In parallel, deep learning techniques have gained traction in respiratory disease diagnosis using medical imaging modalities such as chest X-rays~\cite{bhosale2023puldi}, CT scans~\cite{shuvo2023automated}. While these methods have improved diagnostic accuracy, their applicability is limited in resource-constrained settings due to factors like radiation exposure, high costs, and restricted access to imaging equipment. Unlike imaging-based approaches, lung sound recordings can be obtained in a cost-effective and non-invasive manner. However, LS recordings, particularly in real-world environments, are invariably subject to various interferences. These include heart sounds, microphone contact noise, muscle contractions, noises from other medical equipment, conversations, mobile phones, and assorted ambient disturbances. Such interference can undermine the precision of computer-based LS analyses, creating a risk of misdiagnoses or missed diagnoses. As a result, noise reduction or denoising becomes a critical component in LS signal processing in electronic stethoscopes. Nevertheless, this task is challenging. The noise is unpredictable and dynamic and can differ based on an individual's physical condition. A primary obstacle lies in the temporal and spectral overlap between the essential LS signals and the intrusive noise sources~\cite{pouyani2022lung,yang2023cardiopulmonary,baharanchi2022noise}.
\begin{figure*}
    \centering
    \includegraphics[width=0.9\textwidth]{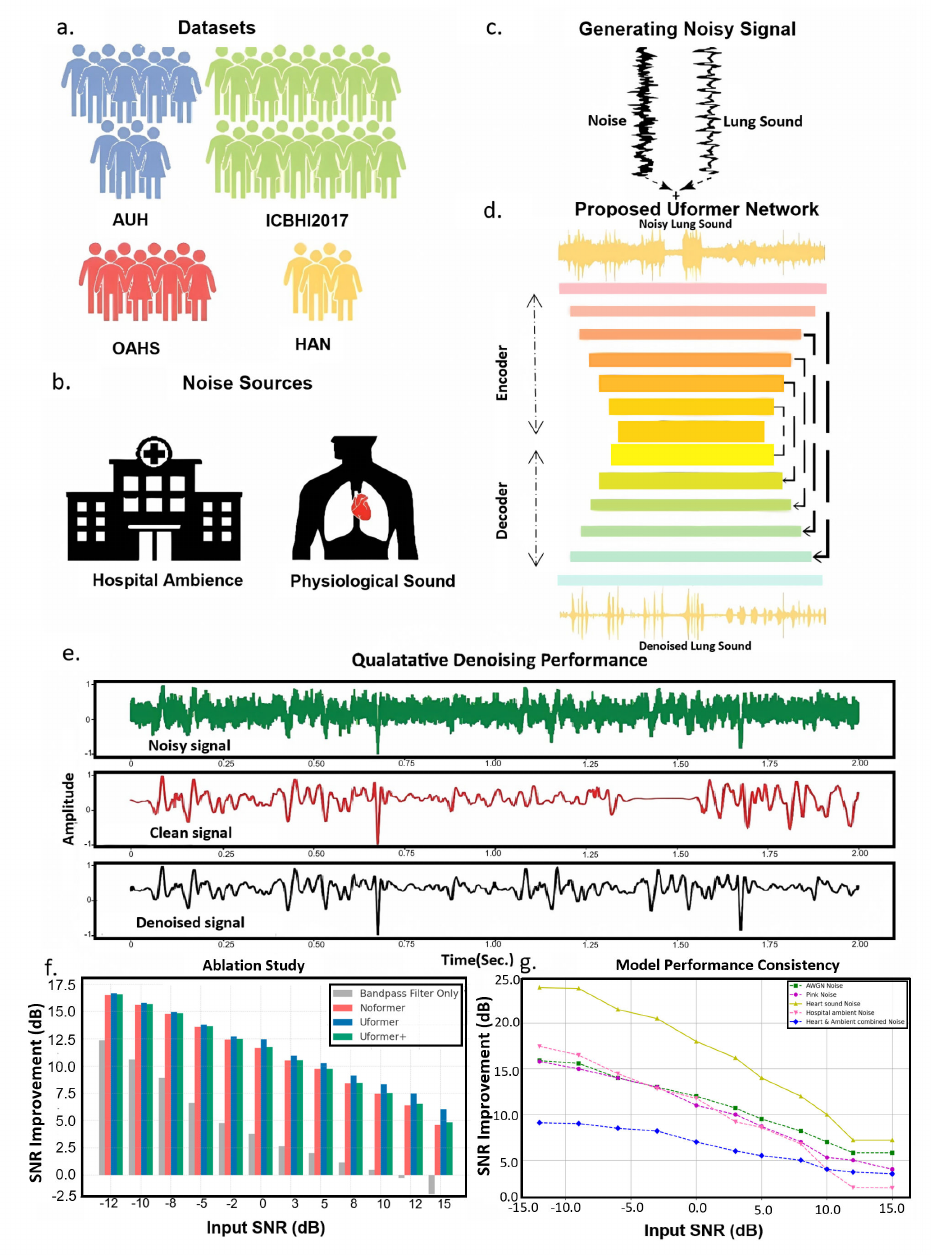}
    \caption{\textbf{(a)} The schematic represents datasets of different signals (LS, HS, and hospital noise) utilized in this research \textbf{(b)} The real world noise sources \textbf{(c)} Generation of input noisy signals \textbf{(d)} Schematic of the proposed Uformer model \textbf{(e)} A qualitative analysis of proposed model: Noisy signal, clean signal, and denoised signal \textbf{(f)} Ablation study: Comparisons of SNR improvement of different variations of the proposed architecture (Uformer+, Uformer,  and Noformer) for LS corrupted with WGN noise \textbf{(g)} Performance of the proposed Uformer model for different types of noise in terms of SNR improvement.}
    \label{fig:GA}
\end{figure*}

\begin{algorithm}[t]
\caption{Generation of Noisy Lung Sound with Heart and Ambient Noise}
\label{algo:mix}

\SetKwInOut{Input}{Input}
\SetKwInOut{Output}{Output}

\Input{$\mathbb{L}_{c,8k,1s}$: Clean lung sound signals sampled at 8kHz, 1 second duration}
\Output{Noisy lung sound signals saved as \texttt{.wav} files}

$SNR_{Desired} \gets [-12, -10, -8, -5, -2, 0, 3, 5, 8, 10, 12, 15]$
$noiseTypes \gets [\texttt{"heart"}, \texttt{"ambient"}]$

\ForEach{$\mathbb{S}_i$ in $\mathbb{L}_{c,8k,1s}$}{
    \ForEach{$SNR_j$ in $SNR_{Desired}$}{
        \ForEach{$type$ in $noiseTypes$}{
            $\mathbb{N} \gets \texttt{GetNoise}(type)$
            $P_{signal} \gets \frac{1}{|\mathbb{S}_i|} \sum \mathbb{S}_i^2$ \tcp*{Signal power}
            $P_{noise} \gets \frac{1}{|\mathbb{N}|} \sum \mathbb{N}^2$ \tcp*{Noise power}
            $\mathcal{SF} \gets \frac{P_{signal}}{P_{noise}} \cdot 10^{-SNR_j / 10}$ \tcp*{Scaling factor for desired SNR}
            $\mathbb{N}_{scaled} \gets \mathbb{N} \cdot \sqrt{\mathcal{SF}}$
            $\mathbb{S}_{Noisy_i} \gets \mathbb{S}_i + \mathbb{N}_{scaled}$
            \tcp*{Write $\mathbb{S}_{Noisy_i}$ to a .wav file labeled with $SNR_j$ and $type$}
        }
    }
}
\end{algorithm}

LS signal denoising methodologies can be primarily divided into classical model-based and learning-based denoising methods. One prevalent technique within the classical model-based category for LS denoising is wavelet-based thresholding methods. These have exhibited commendable performance, especially when the noise-free signal is missing~\cite{singh2020comparative,meng2019kind}. \cite{meng2019kind} integrated FIR band-pass, modified wavelet, and adaptive filters to segregate ambient noise and heart sounds from LS signals. Meanwhile, an alternate approach documented in \cite{syahputra2017noise} employed the Daubechies wavelet, specifically with four coefficients, to eliminate noise. A notable drawback of traditional wavelet-based methods is the introduction of artifacts, such as Gibbs oscillations and noise spikes, particularly around signal discontinuities~\cite{ding2015artifact}. Empirical Wiener filtering was integrated within the wavelet transform domain to alleviate these constraints. However, this technique needs the use of dual wavelet transform bases, and the specific combinations of wavelet bases influence its denoising effectiveness used \cite{ghael1997improved,gallaire1998wavelet,choi1998analysis,nikolaev2000wavelet}. In a separate study \cite{mondal2011reduction}, Empirical Mode Decomposition (EMD) was employed to dissect the LS signal into its constituent components. The objective was to suppress the intrusion of heart sounds into lung sounds, and the results were promising across both frequency and time-frequency domains. Another method \cite{molaie2014chaotic} harnessed local projection to diminish acoustic noise commonly found in hospital surroundings. The research study in \cite{haider2021respiratory} combined EMD with Hurst Analysis and spectral subtraction to filter out high-frequency noise disturbances from LS signals across various settings. \cite{haider2018savitzky} opted for the Savitzky-Golay filter to neutralize varying intensities of Gaussian noise evident in the LS signal. Recent innovations have steered towards adaptive techniques for the dynamic elimination of noise from lung sounds. In~\cite{emmanouilidou2015adaptive}, the authors introduced an automated multiband method to eliminate clinical noise from the LS signal. Interferences from heart sounds, which often have spectral overlap with lung sounds, were tackled in \cite{al2018performance} with the formulation of a Normalized Least-Mean-Square (NLMS) method. This ensured an adaptive distinction between the two sound types.

Transitioning to the realm of artificial intelligence, the integration of deep learning is becoming increasingly prevalent in denoising various physiological signals, including electroencephalogram (EEG)~\cite{yin2023gan,an2022auto}, electrocardiogram (ECG)~\cite{singh2022attention,wang2022ecg}, and phonocardiogram (PCG)~\cite{ali2023end,chandel2022stacked}
. In this context, research works \cite{pouyani2022lung} and \cite{yang2023cardiopulmonary} incorporated deep learning frameworks to denoise LS signals. The study \cite{pouyani2022lung} merged wavelet transform with an artificial neural network (DWT-ANN) for this purpose. Their outcomes reflected that the DWT-ANN outperformed the standalone wavelet transform method. Their methodology utilized synthetic Gaussian white noise at varying signal-to-noise ratios (SNRs) to validate their algorithm. The artificial nature of this noise does not mimic the intricate and unpredictable nature of noise in real-world clinical settings. Similarly, \cite{yang2023cardiopulmonary}  unveiled a method featuring a two-stage process that integrates adaptive noise cancellation (ANC) with deep neural networks (DNNs) to filter out ambient noises and retain the functional signal components. Their approach attempted to simulate real-world scenarios more closely by introducing environmental and hospital noises to clean lung sounds. Nevertheless, the issue of heart sounds interfering with lung sounds due to their anatomical proximity is a challenging real-world problem that was not fully addressed. 
A pivotal limitation of both methodologies is the absence of an end-to-end approach. By combining signal processing with deep learning methodologies, they inadvertently introduced latency in real-time signal processing.

Given the limitations and challenges in existing lung sound denoising methods, we propose an end-to-end self-attention-enabled “Uformer” architecture designed to effectively denoise lung sounds. Our model processes raw noisy lung sound recordings directly and outputs clean denoised signals within a unified deep learning framework, eliminating the need for handcrafted feature extraction or multiple separate processing stages. This holistic approach leverages long-range contextual information from the encoder and integrates it with spatial specifics during the decoding process, resulting in the generation of precise denoised outputs by the decoder. 
The performance of the proposed model is rigorously assessed using a diverse set of challenging real-world noise types, with a particular emphasis on relevant noise like heart sounds. A complete flow diagram of our study is shown in Fig.~\ref{fig:GA}.

The major contributions of our work are as follows: 

\begin{itemize}
\item 
We introduced a novel "Uformer" model tailored specifically for the efficient denoising of lung sounds. This architecture seamlessly combines convolutional and transformer structures in a dual-encoder-decoder framework. 
%This unique combination results in a robust end-to-end denoising performance, surpassing contemporary techniques and underscoring the pronounced superiority of the Uformer.

\item 
We integrated a multi-head attention mechanism into the encoder to enhance the model's ability to handle non-deterministic and overlapping noise in lung sounds. This mechanism captures long-range dependencies, proving essential for addressing complex noise scenarios. 
%The multi-head transformer design improves the encoding of distinct lung sounds by incorporating latent features extracted by the CNN encoder. This comprehensive approach fuses contextual information from the encoder with spatial specifics, yielding precise denoised outputs.
\item 
Conducted experimental validation by infusing lung sounds with diverse noisy conditions, including WGN, pink noise, heart sound noise, and real-world heart and hospital sounds to simulate challenging and varied scenarios. The denoiser's performance was extensively evaluated across a spectrum of signal-to-noise ratios (SNRs) during both the training and testing phases. Moreover, the experimentation spanned a wide range of datasets for training and testing, affirming the generalizability and robustness of the proposed Uformer architecture.

% \item 
% An ablation study was conducted, a systematic approach aimed at refining and ensuring the denoising resilience and effectiveness of the proposed model by identifying the optimal network architecture. 

\end{itemize}
The structure of this paper is as follows: In Section~\ref{sec2}, we present a comprehensive exploration of data resources and an overview of the noisy signal generation scheme. Section~\ref{sec3} provides an in-depth explanation of our proposed method and details of the experimental setup and the evaluation metrics employed. Section~\ref{sec4} describes the experimental details, and both qualitative and quantitative analyses of the results obtained are presented in Section~\ref{sec5}. In Section~\ref{sec6} we describe the shortcomings of this work. Finally, in Section~\ref{sec7}, we summarize our findings and conclude with remarks emphasizing future directions.

\section{Materials and Methods}\label{sec2}

\subsection{Dataset Description}\label{sec2.1}
\begin{figure*}
    \centering
    \renewcommand{\arraystretch}{1}
    \includegraphics[width=0.9\textwidth]{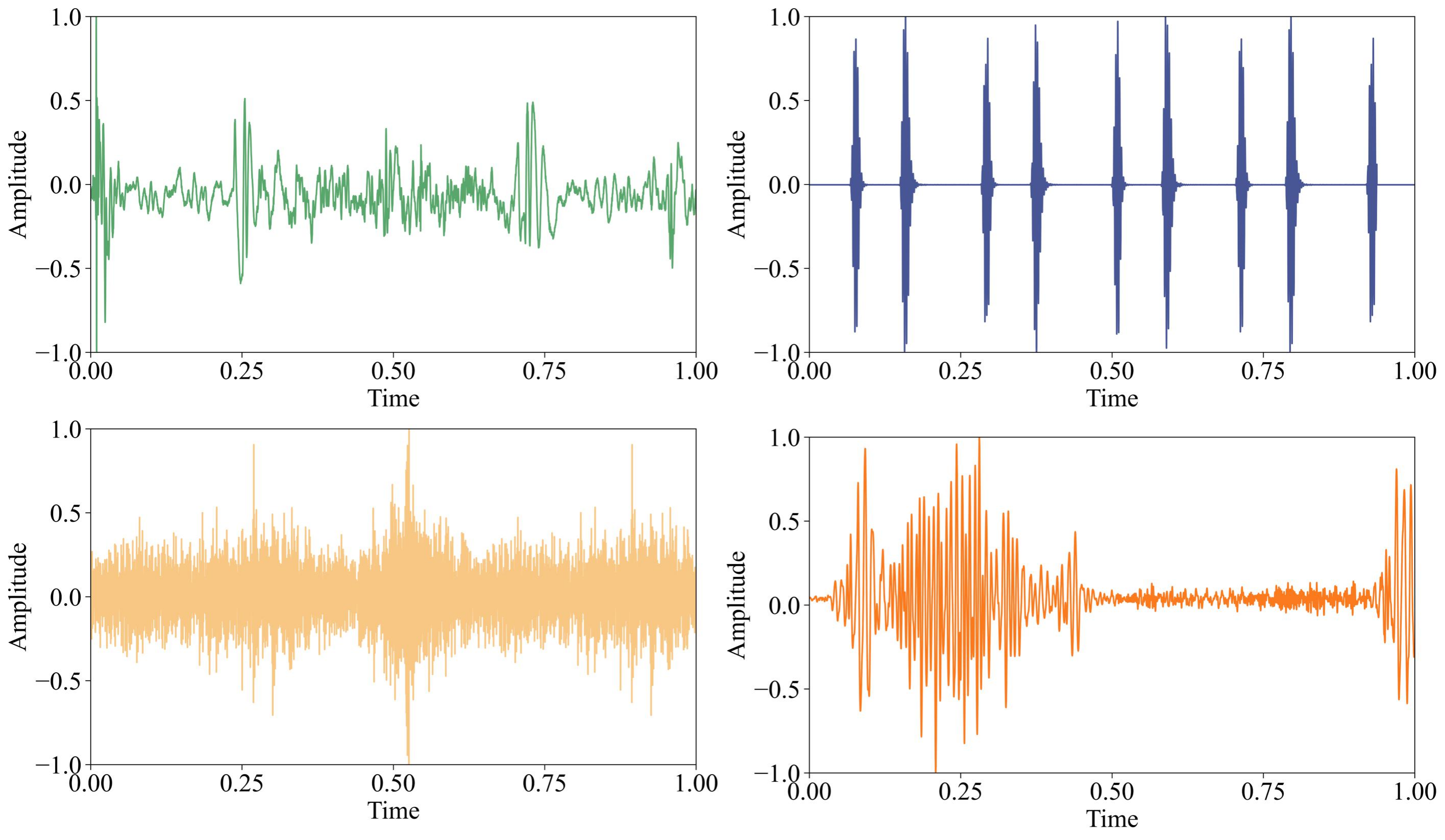}
    \caption{Representative signal of (a) Lung sound, (b) Heart sound signal of OAHS, (c) Hospital ambient sound signal of HAN and (d) Hospital \& Heart sound combined noise signal }
    \label{fig:sample}
\end{figure*}
In this study, two different LS datasets were used(see the class distribution of the Lung sound datasets in Table.~\ref{tab:dataset-class-distribution}). The International Conference on Biomedical Health Informatics (ICBHI) 2017 dataset~\cite{rocha2018alpha} was considered as the primary LS dataset. The secondary independent test LS dataset was the Abdullah University Hospital (AUH) dataset~\cite{fraiwan2021dataset}. The heart sounds and hospital ambient sounds were utilized as noise and collected from the Open Access Heart Sound (OAHS) and Hospital Ambient Noise (HAN) datasets. A brief description of the datasets is discussed in this subsection. 

\begin{table}
\caption{Distribution of respiratory condition classes across ICBHI and AUH datasets}\label{tab:dataset-class-distribution}

\centering
    \renewcommand{\arraystretch}{1.4}
    \resizebox{\linewidth}{!}{%

% \begin{table}[ht]
% \centering
% \caption{Class distribution of lung sound datasets used in this study}
% \label{tab:dataset-class-distribution}
% \begingroup
% \setlength{\tabcolsep}{12pt}
% \renewcommand{\arraystretch}{1.3}
\begin{tabular}{l| c| c}
\hline
\hline
\textbf{Class} & \textbf{ICBHI Dataset} & \textbf{AUH Dataset} \\
\hline
\hline
Pneumonia        & 37  & 5  \\
Bronchiectasis   & 16  & -- \\
COPD             & 793 & 9  \\
Healthy / Normal & 35  & 35 \\
URTI             & 23  & -- \\
Bronchiolitis    & 13  & 3  \\
Asthma           & --  & 32 \\
Heart Failure    & --  & 21 \\
Lung Fibrosis    & --  & 5  \\
Pleural Effusion & --  & 2  \\
\hline
\hline
\end{tabular}}

\end{table}

\subsubsection{ICBHI 2017 Dataset}\label{sec2.1.1}
The International Conference on Biomedical Health Informatics (ICBHI), published a benchmark dataset of lung auscultation sounds for public use in 2017. This dataset was collected by two research groups from Greece and Portugal, containing a total recording of 5.5 hours, sampled at 4KHz, 14KHz, and 44.1KHz with annotated respiratory cycles from 126 patients with 920 recordings~\cite{rocha2018alpha}. The data length of each signal varies from 10sec to 90sec. The dataset includes seven classes, namely, Upper Respiratory Tract Infection, Lower Respiratory Tract Infection, Chronic Obstructive Pulmonary Disease, Bronchiolitis, Bronchiectasis, Pneumonia, and Asthma. A detailed dataset description can be found in \cite{rocha2018alpha}. In this work, we have utilized 80\% of the data for training and 20\% of the data for testing our models, and among the training set, 10\% of the data were set aside for validation.
\subsubsection{AUH Lung Sound Dataset}\label{sec2.1.2}
The AUH dataset contains respiratory signals from 112 Middle Eastern subjects (77 unhealthy and 35 healthy) aged 21-90, collected from King Abdullah University Hospital, Ramtha, Jordan. Each recording is 5-30 seconds long. The dataset has respiratory signals of eight different lung conditions: asthma, pneumonia, BRON, COPD, heart failure, lung fibrosis, pleural effusion, and normal. A single-channel stethoscope was used to record the LS signals. All signals were collected at a sampling rate of 4 KHz using a 16-bit quantizer. The detailed description of the dataset can be found in~\cite{fraiwan2021dataset}. 
\subsubsection{Open Access Heart Sound Dataset}\label{sec2.1.3}
In the present work, the OAHS dataset encompassed five distinct categories: Mitral Regurgitation (MR), Aortic Stenosis (AS), Mitral Stenosis (MS), Mitral Valve Prolapse (MVP), and Normal (N). The signals were collected from diverse sources, including websites, CDs, and books~\cite{yaseen2018classification}. The dataset comprises a total of 1,000 audio recordings, with each entry containing three complete cardiac cycles of clear heart sounds sampled at $8$kHz as single-channel signals. Furthermore, each category is represented by $200$ individual recordings. The processes of sampling, editing, and file conversion were facilitated using the Cool Edit software. Further details about the OAHS dataset can be found in~\cite{yaseen2018classification}.
\subsubsection{HAN Dataset}\label{sec2.1.4}
This dataset has been prepared using a non-copyrighted video from YouTube of $68$ minutes of a busy hospital where the audio occurrences were recorded from different places such as the waiting room, corridor, etc. The dataset contains $562$ samples, each of which has a duration of 5 seconds. Further details of the dataset can be found in~\cite{ali2023end}.

\subsection{Data Prepossessing}\label{sec2.2}
\subsubsection{Lung Sound Preparation} \label{sec2.2.1}
The frequency spectrum of the LS signals usually resides between 50 Hz to 2500 Hz~\cite{reichert2008analysis}. The LS signals utilized in this work were filtered with a 6th-order Butterworth filter to mimic a ground truth clean signal. In contrast, noise signals were not bandpass-filtered prior to mixing. This design choice reflects real-world conditions where background noise (e.g., ambient sounds, heartbeats, device artifacts) may contain energy across a wide frequency range. Filtering noise before addition would result in an unrealistic simulation and artificially simplify the denoising task. Then, LS signals were resampled at 8KHz to ensure consistency and normalized within the range of [-1-1]~\cite{shuvo2023nrc}. In order to reduce complexity, each signal was converted into multiple segments using a non-overlapping window of 1 second. The steps involving LS preparation are discussed in Algo.~\ref{algo:Lung} where $\mathbb{L}_{c}$ and $\mathbb{L}_{c,8k,1s}$ are unsegmented LS signals resampled segmented LS signals, respectively. $ \mathbb{S}_i$ and $\mathbb{S}_{i_{8k}}$ are single LS signals and single resampled segmented LS signals, respectively.

\subsubsection{Generating Synthetic Noisy Signals}\label{sec2.2}

We initially divided the ICBHI 2017 dataset into training and testing sets with a ratio of 80\% for training and 20\% for testing. Then, to assess the denoiser model's robustness for lung sounds, we introduced various types of noise to the original signals, simulating real-world scenarios and challenges. The noise types used for this evaluation included:

\begin{algorithm}[t]
\caption{Lung Sound Preprocessing and Segmentation}
\label{algo:Lung}

\SetKwInOut{Input}{Input}
\SetKwInOut{Output}{Output}

\Input{$\mathbb{L}_c$: List of clean lung sound recordings}
\Output{$\mathbb{L}_{c,8k,1s}$: List of 1-second segments resampled at 8kHz}

\ForEach{$\mathbb{S}_i$ in $\mathbb{L}_c$}{ 
    $\mathbb{S}_{i_{8k}} \gets \text{Resample}(\mathbb{S}_i, s_r = 8000)$ \tcp*{Resample to 8kHz}
    
    $iter_{int} \gets \text{int}(\text{len}(\mathbb{S}_{i_{8k}}) / s_r)$ \tcp*{Number of full 1s segments}
    
    \If{$iter_{int} \neq 0$}{
        $seg_{start} \gets 0$ \\
        \For{$j \gets 1$ \KwTo $iter_{int}$}{
            $seg_{end} \gets seg_{start} + s_r$ \\
            $\mathbb{S}_{seg} \gets \mathbb{S}_{i_{8k}}[seg_{start} : seg_{end}]$ \\
            Append $\mathbb{S}_{seg}$ to $\mathbb{L}_{c,8k,1s}$ \\
            $seg_{start} \gets seg_{end}$
        }
    }
    \Else{
        $residual_{ratio} \gets \text{len}(\mathbb{S}_{i_{8k}}) / s_r$ \\
        
        \If{$residual_{ratio} \geq 0.5$}{
            \If{$iter_{int} == 0$}{
                $res \gets \mathbb{S}_{i_{8k}} ~\|~ \mathbb{S}_{i_{8k}}$ \tcp*{Concatenate to reach 1s}
            }
            \Else{
                $res \gets \mathbb{S}_{i_{8k}}[seg_{start} : ] ~\|~ \mathbb{S}_{i_{8k}}[seg_{start} : ]$ \tcp*{Stack last part to reach 1s}
            }
            Append $res[0 : s_r]$ to $\mathbb{L}_{c,8k,1s}$
        }
    }
}
\end{algorithm}

\begin{algorithm}[t]
\caption{Generation of Noisy Lung Sound with White Gaussian Noise (WGN)}
\label{algo:AWGN}

\SetKwInOut{Input}{Input}
\SetKwInOut{Output}{Output}

\Input{$\mathbb{L}_{c,8k,1s}$: Clean lung sound signals sampled at 8\,kHz with 1-second duration}
\Output{Noisy lung sound signals $\mathbb{S}_{\text{Noisy}}$ saved as \texttt{.wav} files for each desired SNR}

$\mu_{\text{noise}} \gets 0$ \tcp*{Mean of noise}

\ForEach{$\mathbb{S}_i \in \mathbb{L}_{c,8k,1s}$}{
    $SNR_{\text{Desired}} \gets [-12, -10, -8, -5, -2, 0, 3, 5, 8, 10, 12, 15]$ \tcp*{List of desired SNRs}
    
    \For{$j \gets 1$ \KwTo $|SNR_{\text{Desired}}|$}{
        $SNR_j \gets SNR_{\text{Desired}}[j]$ \\
        
        $\mathbb{P}_{S_i} \gets \frac{1}{|\mathbb{S}_i|} \sum \mathbb{S}_i^2$ \tcp*{Power of clean signal}
        
        $\mathbb{P}_{\text{Noise}} \gets \mathbb{P}_{S_i} \times 10^{-SNR_j / 10}$ \tcp*{Noise power for desired SNR}
        
        $\mathbb{N}_j \gets \eta(\mu_{\text{noise}}, \sqrt{\mathbb{P}_{\text{Noise}}})$ \tcp*{Generate WGN with mean and std. dev.}
        
        $\mathbb{S}_{\text{Noisy}_i} \gets \mathbb{S}_i + \mathbb{N}_j$ \tcp*{Add WGN to clean signal}
        
        \tcp*{Save $\mathbb{S}_{\text{Noisy}_i}$ as a .wav file labeled with $SNR_j$}
    }
}
\end{algorithm}

\begin{algorithm}[t]
\caption{Generation of Noisy Lung Sounds with Pink Noise}
\label{algo:Pink}

\SetKwInOut{Input}{Input}
\SetKwInOut{Output}{Output}

\Input{$\mathbb{L}_{c,8k,1s}$: Clean lung sound signals sampled at 8\,kHz with 1-second duration}
\Output{Noisy lung sound signals $\mathbb{S}_{\text{Noisy}}$ saved as \texttt{.wav} files for each desired SNR}

$\mu_{\text{noise}} \gets 0$ \tcp*{Mean of pink noise (zero-mean)}

\ForEach{$\mathbb{S}_i \in \mathbb{L}_{c,8k,1s}$}{
    $SNR_{\text{Desired}} \gets [-12, -10, -8, -5, -2, 0, 3, 5, 8, 10, 12, 15]$ \tcp*{List of desired SNRs}
    
    \For{$j \gets 1$ \KwTo $|SNR_{\text{Desired}}|$}{
        $SNR_j \gets SNR_{\text{Desired}}[j]$ \\
        
        $\mathbb{N}_{\text{pink}} \gets \texttt{generate\_pink\_noise}(\text{length} = |\mathbb{S}_i|)$ \tcp*{Generate pink noise with PSD $\propto 1/f$}
        
        $\mathbb{P}_{\text{signal}} \gets \frac{1}{|\mathbb{S}_i|} \sum \mathbb{S}_i^2$ \tcp*{Power of clean signal}
        
        $\mathbb{P}_{\text{target}} \gets \mathbb{P}_{\text{signal}} \times 10^{-SNR_j / 10}$ \tcp*{Target noise power for desired SNR}
        
        $\sigma_{\text{pink}} \gets \sqrt{\frac{1}{|\mathbb{N}_{\text{pink}}|} \sum \mathbb{N}_{\text{pink}}^2}$ \tcp*{Std. dev. of original pink noise}
        
        $\mathbb{N}_{\text{scaled}} \gets \mathbb{N}_{\text{pink}} \times \frac{\sqrt{\mathbb{P}_{\text{target}}}}{\sigma_{\text{pink}}}$ \tcp*{Scale pink noise to match target power}
        
        $\mathbb{S}_{\text{Noisy}_i} \gets \mathbb{S}_i + \mathbb{N}_{\text{scaled}}$ \tcp*{Add scaled pink noise to signal}
        
        \tcp*{Save $\mathbb{S}_{\text{Noisy}_i}$ as a .wav file labeled with $SNR_j$}
    }
}
\end{algorithm}

\begin{algorithm}[t]
\caption{GetNoise Function}
\label{algo:func}

\SetKwFunction{GetNoise}{GetNoise}
\SetKwProg{Fn}{Function}{:}{}
\Fn{\GetNoise{$type$}}{
    \If{$type = \texttt{"heart"}$}{
        $\mathbb{N} \gets$ Randomly select a heart sound \\
        $\mathbb{N}_{8k} \gets$ Resample $\mathbb{N}$ at 8kHz \\
        \KwRet $\mathbb{N}_{8k}$
    }
    \ElseIf{$type = \texttt{"ambient"}$}{
        $\mathbb{N}_{heart} \gets$ Randomly select a heart sound \\
        $\mathbb{N}_{heart,8k} \gets$ Resample $\mathbb{N}_{heart}$ at 8kHz \\
        
        $\mathbb{N}_{hospital} \gets$ Randomly select a hospital sound \\
        $\mathbb{N}_{hospital,8k} \gets$ Resample $\mathbb{N}_{hospital}$ at 8kHz \\
        
        \KwRet $0.7 \cdot \mathbb{N}_{heart,8k} + 0.3 \cdot \mathbb{N}_{hospital,8k}$
    }
}
\end{algorithm}

\textbf{White Noise and Pink Noise: }
In contrast to the controlled and reduced noise levels typically encountered in acoustic labs, the real-world environment of operating rooms introduces significant challenges. High noise levels originate from surgical devices, ventilation machines, conversations, alarms, and other sources, which make analyzing lung sounds more complex. Additional noise can be introduced by the sliding motions of stethoscope diaphragms, resulting from physician adjustments or patient agitation, further complicating the diagnostic process.

In this context, noise elimination techniques become essential to mitigate the detrimental impact of such noise. Environmental noises are often considered colored noises, which means their power spectral density varies with frequency rather than being flat. Therefore, we approximated background noise as additive white Gaussian noise (WGN) and pink noise when capturing lung sounds. WGN and pink noise can also approximate various real-world noises, including fan noise, air conditioner noise, and other static sounds from medical devices \cite{pouyani2022lung}. WGN is characterized by a uniform distribution of noise power across all frequencies (i.e., a flat power spectrum) and is commonly used in deep learning-based classification and denoising tasks \cite{pouyani2022lung, alam2023rf, el2020blind}. In contrast, pink noise is a type of colored noise whose power spectral density decreases inversely with frequency, following approximately a \(1/f\) relationship. This means pink noise contains more power in lower frequencies, closely resembling many real-world noise sources. Consequently, in our evaluation, we considered both WGN and pink noise to assess the denoiser model’s robustness against synthetic and more naturalistic noise environments.

We contaminated LS signals from the ICBHI 2017 Dataset to WGN and Pink Noise at various SNRs during the model's training to evaluate our enhancement algorithm comprehensively. These SNRs encompassed a wide range, including -10 dB, -5 dB, 0 dB, 5 dB, 10 dB, and 15 dB.

In addition to these SNRs, we sought to assess the model's adaptability to the unseen noise levels not used in the training phase. Therefore, we conducted an evaluation using LS signals at SNRs of -12 dB, -8 dB, -2 dB, 3 dB, 8 dB, and 12 dB, further expanding the scope of our assessment.

To further evaluate the model's robustness and its ability to adapt to noise introduced from external datasets, we conducted tests using the AUH Lung Sound Dataset. This evaluation involved the introduction of WGN at SNRs of -12 dB, -10 dB, -8 dB, -5 dB, -2 dB, 0 dB, 3 dB, 5 dB, 8 dB, 10 dB, 12 dB, and 15 dB. This comprehensive evaluation approach ensured that our denoiser model could effectively handle a wide array of noise levels, both from its training and external independent test datasets, making it a versatile and reliable solution for addressing noisy medical environments. The step-by-step process of inducing WGN and pink noise in LS is illustrated in Algo. \ref{algo:AWGN} and \ref{algo:Pink}.

\textbf{Heart Sound Noise and Hospital Ambient Noise:} 
Though denoiser models are typically trained and evaluated on synthetic noises, real-world noises are often more complex, including spatially-correlated, spatially-variant, device-dependent, and signal-dependent components \cite{zhou2020awgn}. Therefore, evaluating models solely based on synthetic noise may not suffice. In the case of lung sounds, the primary source of signal deterioration is the presence of heart sounds.

Heart sounds, specifically the first (S1) and the second (S2) components of normal heart sound recordings, fall within the frequency range of 20 Hz to 150 Hz \cite{ren2018novel}. This frequency range overlaps with typical lung sounds, which may vary between 50 Hz and 2500 Hz. Distinguishing pure respiratory signals from heart sounds can be challenging, as both signals originate from anatomical sites close to each other \cite{nersisson2017heart}. In addition to heart sound noise, hospital ambient noise contributes to the degradation of respiratory signals. Sources of hospital ambient noise include human conversation, intestinal activity, stethoscope movement, and sensor variations \cite{liu2016open}.

To address the challenge posed by real-world noises, we trained and evaluated our denoiser model on both LS signals with heart sound noise and combination of heart and hospital noise. We used the Open-Access Heart Sound (OAHS) dataset for heart sound noise and the Hospital Ambient Noise (HAN) dataset for hospital ambient noise. 
During the training phase, we introduced these noise types at various signal-to-noise ratios (SNRs) into the LS signals from the ICBHI 2017 Dataset. The SNRs considered for training were -10 dB, -5 dB, 0 dB, 5 dB, 10 dB, and 15 dB. Additionally, our model's adaptability to unseen noise levels was thoroughly assessed by subjecting it to LS signals at SNRs of -12 dB, -8 dB, -2 dB, 3 dB, 8 dB, and 12 dB. We also conducted tests using the AUH Lung Sound Dataset for this noise types. Algo. \ref{algo:func} and \ref{algo:mix} are delineates the step-by-step process of inducing heart sound and ambient noise.
\begin{table*}
\caption{Specifications of training and testing datasets with different noise levels}\label{dataset}

\centering
    \renewcommand{\arraystretch}{1.4}
    \resizebox{\linewidth}{!}{%
\begin{tabular}{c|c|c|c|c|c}  
\hline
\hline
Split & Dataset Name & Name & Noise Type & Noise level(dB)
& \parbox[c]{2.5cm}{\centering No. of samples (1s segment)} \\
\hline
\hline
\multirow{4}{*}{\parbox[c]{2.5cm}{\centering Training(90\%) Validation(10\%) }} & \multirow{4}{*}{\parbox[c]{2.5cm}{\centering ICBHI-2017 (731 recordings)}} & ICBHI-1 & WGN & \multirow{4}{*}{[-10,-5,0,5,10,15]}
& \multirow{4}{*}{92556} \\
 &  & ICBHI-2 & Pink Noise & 
&  \\
 &  & ICBHI-3 & Heart Sound Noise & 
&  \\
 &  & ICBHI-4 &  Heart+Hospital Noise & 
&  \\
 \hline
 \hline
\multirow{8}{*}{Testing } & \multirow{4}{*}{\parbox[c]{2.5cm}{\centering ICBHI-2017 (186 recordings)}} &
ICBHI\_inf-1 & WGN &  \multirow{8}{*}{\parbox[c]{3cm}{\centering [-10,-5,0,5,10,15], [-12,-8,-2,3,8,12] (unseen noise level)} } & \multirow{4}{*}{25410} \\
 &  & ICBHI\_inf-2 & Pink Noise & 
&  \\
 &  & ICBHI\_inf-3 & Heart Sound Noise & 
&  \\
 &  & ICBHI\_inf-4 & Heart + Hospital Noise & 
&  \\
\cline{2-4}
\cline{6-6}
 & \multirow{4}{*}{\parbox[c]{2.5cm}{\centering AUH Lung Sound Dataset (55 recordings)}} &
AUH-1 & WGN &  & \multirow{4}{*}{9745} \\
 &  & AUH-2 & Pink Noise & 
&  \\
 &  & AUH-3 & Heart Sound Noise & 
&  \\
 &  & AUH-4 & Heart + Hospital Noise & 
&  \\

\hline
\hline
\end{tabular}}

\end{table*}

A brief overview of the specifications of the training and testing datasets demonstrated in Table~\ref{dataset} .

\section{Uformer Architecture}\label{sec3}

\begin{figure*}[t!]
    \centering
    \includegraphics[width=0.98\textwidth]{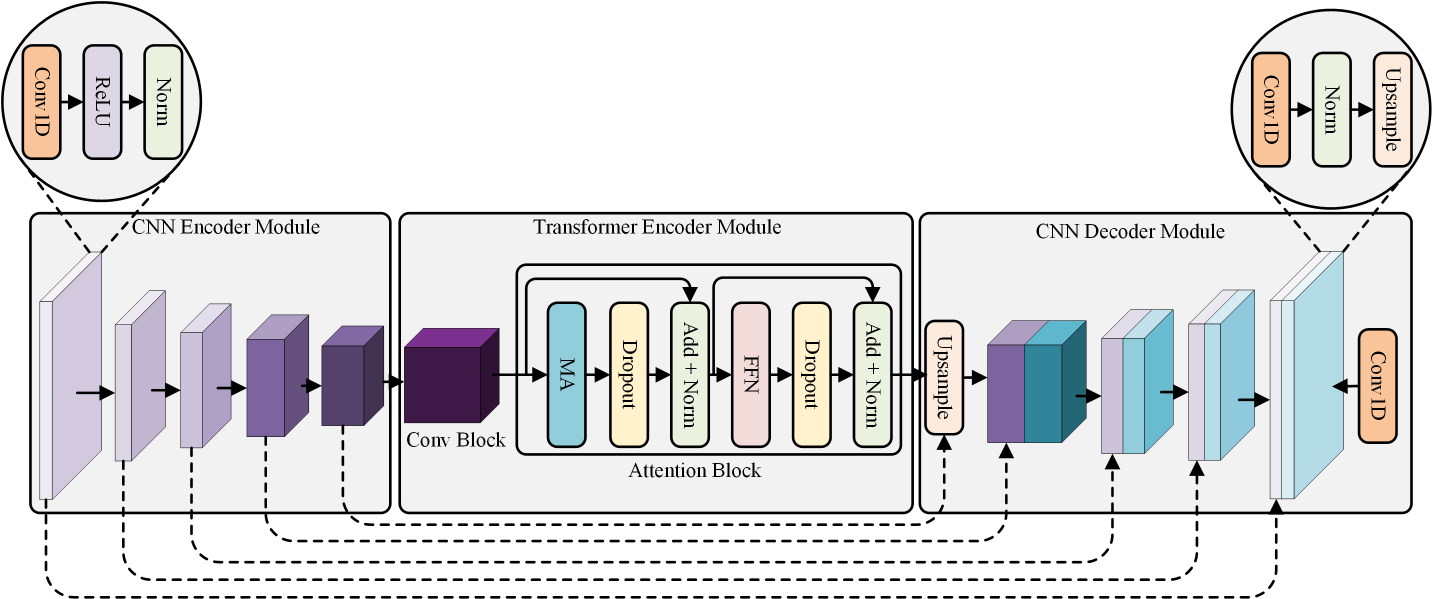}

    \caption{{The proposed Uformer network}}
    \label{fig:Netwrok}
\end{figure*}

Our proposed Uformer model is designed to perform denoising on noisy LS. The architecture, tailored for temporal domain signals, integrates three principal components: a CNN encoder module for latent features extraction, a transformer encoder module to refine the encoding of distinctive lung sounds and capture fine long-range dependencies, and a CNN decoder to produce enhanced denoised signals (see detailed visual description in Fig.~\ref{fig:Netwrok}). Detailed insights into each module are given in the following subsections.

\subsection{Input Formulation}\label{sec2.4.0}
The noisy LS signals are represented as a tensor \( x \in \mathbb{R}^{N \times C} \) where \( C = 1 \) corresponds to the number of channels, and \( N = 8000 \) represents the count of noisy samples. The entire dataset can be articulated as \( X = \{x_1, x_2, \ldots, x_N\} \).

\subsection{CNN Encoder Module}\label{sec2.4.1}
The CNN encoder module symbolized as \(E^{CNN}\), draws inspiration from the U-Net design. It comprises five recurring blocks, \(E_b^{CNN}\). Each block consists of a Conv1d-ReLU pair, utilizing a kernel size of 31 and a stride of 2. Following these operations, a Batch-normalization layer is applied. For a given input signal \(x\), each CNN encoder block outputs 
\begin{equation}
f_{a,n} = \{f_{a1,n1}, f_{a2,n2}, \ldots, f_{a5,n5}\}
\end{equation}, where \(f_{a,n} \in \mathbb{R}^{N_a \times C_a}\) denotes the feature map at each block level. Here, \('a'\) ranges from \(1\) to \(5\), and \(n \in\{16, 32, 32, 64, 64\}\) signifies the filter count. \(N_a\) and \(C_a\) represent the feature map size and channel number at level \(a\), respectively. The final feature map \(f_{a5,n5} =z'\) serves as the input for the transformer encoder module.

\subsection{Transformer Encoder Module}\label{sec2.4.2}
The transformer encoder module, denoted as \(E^{Trans}\), is structured into two blocks: the convolutional block, \(E^{Trans}_c\), and the attention block, \(E^{Trans}_{att}\). To ensure minimal memory usage, \(E^{Trans}_c\) further compresses the output feature map \(z^{'}\) from \(E^{CNN}\).  The \(E^{Trans}_c\) block incorporates Conv1d-ReLU operations with 128 filters, using a stride of 2, succeeded by Batch-normalization layers. The \(E^{Trans}_{att}\) block comprises two sub-layers: the first employs a multi-head self-attention mechanism (\(\mathcal{MA}\)), and the second integrates a position-wise, fully connected feed-forward network.

Given the current representations \(\mathbf{m}_t = \{ \mathbf{m}_1, \mathbf{m}_2, \ldots , \mathbf{m}_N \}\) from the convolutional block, the refined representation \(\mathbf{n}_t\) is determined as:

\begin{align}
n_t &= \mathcal{MA}(m_1, m_2, \ldots, m_N) \nonumber\\
&= \Biggl[\mathcal{A}_1(Q, K, V) (m_t) \oplus \mathcal{A}_2(Q, K, V) (m_t)  \nonumber\\
&                    \mathcal{A}_i(Q, K, V) (m_t)\Biggl]W_o
\end{align}

The attention function \(\mathcal{A}\) is defined as:
\begin{equation}
\mathcal{A}(\mathbf{Q}, \mathbf{K}, \mathbf{V}) = \mathcal{SM}\left(\frac{\mathbf{W_Q} \mathbf{W_K}^T}{\sqrt{d_k}}\right) \mathbf{W_V}
\end{equation}

Where \(\mathcal{SM}\) signifies softmax activation, and \(\mathbf{W_Q}\), \(\mathbf{W_K}\), and \(\mathbf{W_V}\) are weight matrices for queries, keys, and values respectively. Dimensions of queries, keys\(d_k\) and values is 25.

The sub-layer's output is then calculated as:
\begin{equation}
g = \mathcal{LN}(\mathbf{m}_t + \mathcal{MA}(\mathbf{m}_t))
\end{equation}

Where \(\mathcal{LN}\) symbolizes the layer normalization layer.

The second sub-layer employs two dense layers (\(\mathcal{D}\)) with ReLU activation in between, each containing 128 nodes. Residual connections link the two sub-layers, succeeded by layer normalization. The sub-layer's output is represented as:

\begin{equation}
h = \mathcal{LN}\Biggl(\mathcal{D}(\mathbf{ReLU}(\mathcal{D}(\mathbf{g}))) + \mathbf{g}\Biggl)
\end{equation}

Subsequently, the feature map undergoes upsampling (factor of 2) and is concatenated with \(\mathbf{f}_{a5,n5}\) to yield the final encoded feature map \(\mathbf{z} = E^{\text{Trans}}(z')\).

\subsection{CNN Decoder Module}\label{sec2.4.3}
The CNN decoder module, symbolized as \(D^{CNN}\), contains 4 blocks, termed \(D_b^{CNN}\) blocks, where b=1 to 4. Each block integrates pairs of Conv1d-ReLU operations using a kernel size of 31 and a stride of 1. These operations are followed by batch normalization and Upsample layers with a scale factor of 2.

The encoder-derived feature map, \(z\), serves as the decoder input. It undergoes upsampling and passes through the decoder blocks to reconstruct high-resolution signals. Each feature map of a CNN decoder block is designated as 
\begin{equation}
d_{a,n} = \{d_{a1,n1}, d_{a2,n2},...,d_{a4,n4}\}.\nonumber
\end{equation}
where $a$ denotes the block level (from 1 to 4), and $n$ relates to the filter count in each convolutional layer, with \(n \in \{64, 32, 32, 32\}\). Skip connections bridge the encoder and decoder to ensure the retention of intricate details, significantly enhancing denoised LS reconstruction. The combined feature maps, labeled as \(k_{a,n}\), are expressed as:
\begin{equation}
k_{a,n} = h_{a,n} \oplus f_{5-a,n} 
\end{equation}
where \(\oplus\) represents concatenation of two vectors.

\subsection{Output Formulation}\label{sec2.4.4}
The final feature map from the decoder module, \(k_{4,32}\), traverses a concluding Conv1d-Tanh layer with a single filter. This process yields the ultimate denoised signal $\hat{y} = \{\hat{y}_1, \hat{y}_2, ..., \hat{y}_N\}$
%\(\hat{y} = \{\hat{y}_1, \hat{y}_2, ..., \hat{y}_N\}\), 
where \(\hat{y} \in \mathbb{R}^{N \times 1}\) and \(N = 8000\).

\par Encoder–decoder architectures, while powerful, can sometimes suffer from challenges such as information loss during downsampling, mode collapse, and a lack of alignment between input and output. To mitigate these, the proposed Uformer model incorporates skip connections between corresponding encoder and decoder layers, ensuring retention of fine-grained temporal features. Additionally, the use of a transformer-based attention block enables the model to capture global contextual dependencies, enhancing representation quality and reducing the risk of feature redundancy. The final convolutional layer with \texttt{tanh} activation enforces consistent output dimensionality, ensuring alignment with the input. These design choices are further validated through superior empirical performance across multiple noise conditions and datasets.
\begin{table*}[!ht]
    \centering
\caption{Simulation parameters of the proposed Uformer model}
\label{tab:uformer_params}
    \renewcommand{\arraystretch}{1}
\resizebox{\linewidth}{!}{

% \begin{table*}[htbp]
% \centering

\begin{tabular}{p{4cm}|p{5cm}|p{7cm}}
\hline
\hline
\textbf{Component} & \textbf{Parameter} & \textbf{Value / Description} \\
\hline
\hline
Input Shape & Segment length & (8000, 1) \\
\hline

\multirow{3}{*}{Encoder Layers} & Conv1D filter sizes & 16, 32, 32, 64, 64, 128 \\
 & Kernel size / Stride & 31 / (1 or 2) \\
 & Activation & ReLU \\
 & Normalization & Batch Normalization \\
\hline

\multirow{4}{*}{Attention Block} & Attention heads (num\_heads) & 8 \\
 & Key dimension (key\_dim) & 25 \\
 & Feedforward hidden units & 128 \\
 & Dropout rate & 0 (disabled) \\
\hline

\multirow{4}{*}{Decoder Layers} & Conv1D filter sizes & 64, 32, 32, 32, 32 \\
 & Kernel size / Stride & 31 / 1 \\
 & Upsampling & UpSampling1D (factor = 2) \\
 & Skip connections & Used (from encoder layers) \\
\hline

Output Layer & Conv1D + Activation & 1 filter, Tanh activation \\
\hline

\multirow{7}{*}{Training Configuration} & Epochs & 500 \\
 & Batch size & 150 \\
 & Optimizer & Adam (learning rate = $10^{-4}$) \\
 & Loss function & Mean Squared Error (MSE) \\
 & Early stopping & Patience = 5 epochs (monitoring val\_loss) \\
 & Model checkpoint & Saves best model weights \\
 & Validation split & 10\% of training data \\
 & Random seed & 111 \\
\hline

\multirow{3}{*}{Framework \& Hardware} & Framework & TensorFlow 2.0 (Keras backend) \\
 & Hardware & Intel i7 CPU, RTX 4090 GPU \\
 & RAM & 128 GB \\
\hline
\hline
\end{tabular}}

\end{table*}

\section{Experimental Setup}\label{sec4}

This section outlines the implementation details, training configuration, and architectural parameters used for the proposed Uformer model. A summary of the main simulation parameters, including the encoder and decoder structure, the settings of attention and the training strategy, is presented in Table~\ref{tab:uformer_params}. This table provides a consolidated overview for reproducibility and clarity.

% \endgroup

\subsection{Loss Function}
The Mean Squared Error (MSE), which we used in our training process, quantifies the dissimilarity between the denoised signal and the ground truth signal. 
This function measures the average squared difference between the corresponding samples of the denoised signal \(\hat{y}\) and the clean input signal \(y\). Minimizing the MSE encourages our model to produce denoised signals that closely match the true, noise-free signal. 
The MSE loss can be expressed as follows :
\begin{equation}
  \mathcal{L}\textsubscript{\textbf{MSE}} = \frac{1}{N} \sum_{i=1}^{N} (x_i - y_i)^2  
\end{equation}
where \(N\) represents the total number of samples in the signals, \(y_i\) is the \(i\)th sample of the clean lung sound signal, \(\hat{y}_i\) is the \(i\)th sample of the denoised LS signal.
\subsection{Hyperparameters}
For training both models, we employ the Adam~\cite{kingma2014adam}, with an initial learning rate (lr) set to $10^{-4}$ and a batch size of $512$. We opt for the Adam optimizer due to its superior efficacy compared to other optimization methods, and its stochastic momentum-based approach is chosen to expedite the model training process~\cite{wilson2017marginal}.

\subsection{Hardware}
Our models have been developed using Keras and TensorFlow 2.0 and trained using an Intel 12th Gen Core i7-12700F 5GHz CPU, 128 GB RAM, and a GeForce RTX 4090 GPU with 24 GB VRAM.
\subsection{Evaluation Metrics}
\subsubsection{Signal-to-Noise Ratio (SNR)}
Signal-to-Noise Ratio (SNR) is a primary metric for evaluating noise reduction performance in LS denoising. It is defined as follows:
\begin{align}
    \label{e1}
    \mathcal{E}_{\textbf{SNR (dB)}} &= 10\log_{10}\left(\frac{P_{\hat{y}}}{P_n}\right) \nonumber \\
    &= 10\log_{10}\left( \frac{\sum_{t=1}^{N}\hat{y}[t]^2}{\sum_{t=1}^{N} (y[t] - \hat{y}[t])^2} \right)
\end{align}
where $\hat{x}$ is the denoised signal and $x$ is the clean reference. This SNR formulation is widely used in denoising literature to evaluate how closely the denoised signal approximates the ground truth\footnote{It is important to note that if the denoiser is disabled (i.e., $\hat{x} = y = x + n$), Equation~\ref{e1} becomes the ratio of the power of the signal-plus-noise to the power of the noise, which differs from the conventional definition of input SNR. Therefore, Equation~\ref{e1} should be interpreted strictly as a performance metric for comparing denoising quality when clean references are available.}.

\subsubsection{Root-Mean-Squared Error (RMSE)}
Root-mean-squared error (RMSE) is commonly used to measure the difference between the intended and actual signals. A lower RMSE indicates a smaller difference. 
The RMSE between the clean and denoised LS recordings is calculated as:
\begin{equation}
    \mathcal{E}\textsubscript{\textbf{RMSE}} = \sqrt{\frac{1}{N}\sum_{t=1}^{N}{({y[t]-\hat{y}[t]})^2}}
\end{equation}
where \(y[t]\) and \(\hat{y}[t]\) denote the clean and denoised LS signals, respectively, and \(N\) is the signal length.

\subsubsection{Short-Time Mean Absolute Error (ST-MAE)}
The Short-Time Mean Absolute Error (ST-MAE) quantifies the average absolute difference between the clean and denoised signals within short overlapping windows. It provides insight into the instantaneous temporal fidelity of the denoising process. A lower ST-MAE indicates better localized accuracy in the denoised output.

The ST-MAE is computed by averaging the absolute error over a window of length \(W\), slid across the signal with step size \(S\), and is defined as:
\begin{equation}
    \mathcal{E}\textsubscript{\textbf{ST-MAE}}(t) = \frac{1}{W} \sum_{i=t}^{t+W-1} \left| y[i] - \hat{y}[i] \right|
\end{equation}
where \(y[i]\) and \(\hat{y}[i]\) represent the clean and denoised lung sound (LS) signals, respectively, \(W\) is the window length, and \(t\) denotes the window's starting index. This metric is computed over all valid window positions and can be visualized as a time series reflecting localized denoising performance.

\subsubsection{Percent Root-Mean-Squared Difference (PRD)}
The Percent Root-Mean-Squared Difference (PRD) metric assesses the recovery efficiency by comparing the input signal to the reconstructed signal. A lower PRD value indicates better reconstruction. 
PRD is calculated as follows:
\begin{equation}
    \mathcal{E}\textsubscript{\textbf{PRD}} = \sqrt{\frac{\sum_{t=1}^{N}(y[t] -\hat{y}[t])^2}{\sum_{t=1}^{N}(y[t])^2}}
\end{equation}
where \(y[t]\) and \(\hat{y}[t]\) represent the clean and denoised LS signals, respectively, and \(N\) is the signal length.
\begin{table*}[!ht]
    \centering
    \caption{Ablation study results comparing the performance of different attention mechanisms}\label{ablation}
    \renewcommand{\arraystretch}{1.2}
    \resizebox{\linewidth}{!}{%
    \begin{tabular}{c|c|c|c|c|c|c|c|c|c}
    \hline \hline
        \multirow{3}{*}{\parbox[c]{1.5cm}{\centering \textbf{Noise Level \\ (in dB)}}} & \multicolumn{9}{c}{\textbf{Model Configurations}} \\ \cline{2-10}
        & \multicolumn{3}{c|}{Uformer+} & \multicolumn{3}{c|}{Uformer} & \multicolumn{3}{c}{Noformer} \\ \cline{2-10}
        & PRED\_SNR & PRD & RMSE & PRED\_SNR & PRD & RMSE & PRED\_SNR & PRD & RMSE \\\hline \hline
-12 & 4.5772 & 0.9750
& 0.0583 & 4.6709 & 0.9710 & 0.0579 & 4.5002 & 0.9770
& 0.0586 \\
        -10 & 5.6953 & 0.9583
& 0.0512 & 5.7871 & 0.9543 & 0.0509 & 5.6226 & 0.9603
& 0.0515 \\ 
        -8 & 6.8450 & 0.9338
& 0.0447 & 6.9504 & 0.9298 & 0.0445 & 6.7811 & 0.9358
& 0.0450 \\ 
        -5 & 8.6335 & 0.8759
& 0.0361 & 8.7927 & 0.8719 & 0.0359 & 8.5871 & 0.8779
& 0.0363 \\ 
        -2 & 10.4977 & 0.7872
& 0.0289 & 10.7145 & 0.7832 & 0.0286 & 10.4346 & 0.7892
& 0.0290 \\ 
        0 & 11.7299 & 0.7113
& 0.0248 & 12.45 & 0.7073 & 0.0245 & 11.6545 & 0.7133
& 0.0249 \\ 
        3 & 13.5375 & 0.5818
& 0.0198 & 13.9523 & 0.5778 & 0.0194 & 13.5156 & 0.5838
& 0.0198 \\ 
        5 & 14.7288 & 0.4941
& 0.0170 & 15.2524 & 0.4901 & 0.0165 & 14.7244 & 0.4961
& 0.0171 \\ 
        8 & 16.4601 & 0.3739
& 0.0137 & 17.1211 & 0.3699 & 0.0131 & 16.4228 & 0.3759
& 0.0138 \\ 
        10 & 17.5372 & 0.3055
& 0.0119 & 18.3306 & 0.3015 & 0.0113 & 17.4549 & 0.3075
& 0.0121 \\ 
        12 & 18.5405 & 0.2476
& 0.0105 & 19.4779 & 0.2436 & 0.0098 & 18.3890 & 0.2496
& 0.0107 \\ 
        15 & 19.8325 & 0.1791& 0.0088 & 21.0526 & 0.1751 & 0.0080 & 19.5921 & 0.1811& 0.0092 \\ \hline\hline 
    \end{tabular}
    }
\end{table*}

\begin{figure}[!ht]
    \centering
    \renewcommand{\arraystretch}{1}
    \includegraphics[width=1\linewidth]{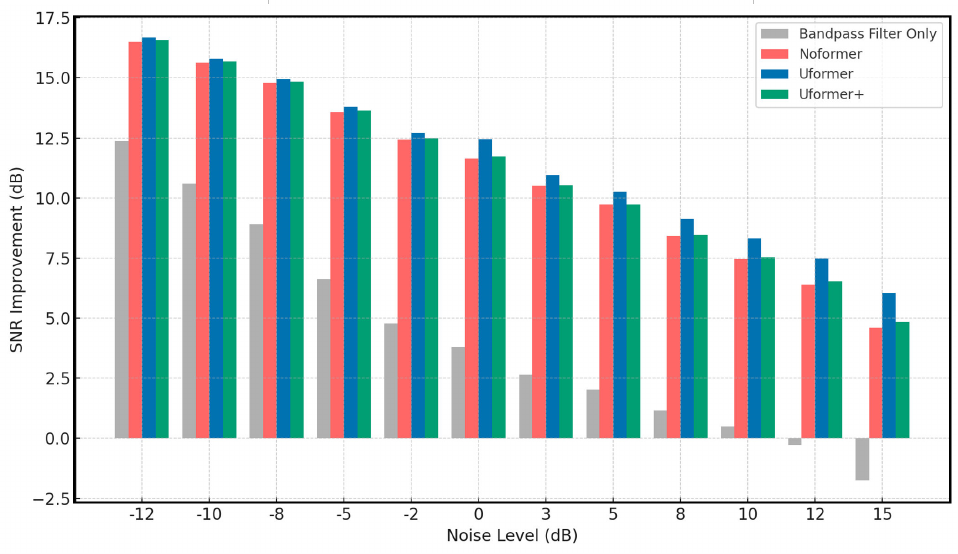}

    \caption{Comparison of SNR improvement across noise levels for different attention configurations and bandpass baseline}

    \label{fig:SNR_Im}
\end{figure}

\begin{figure*}
    \centering
    \renewcommand{\arraystretch}{1}
    \includegraphics[width=1\textwidth]{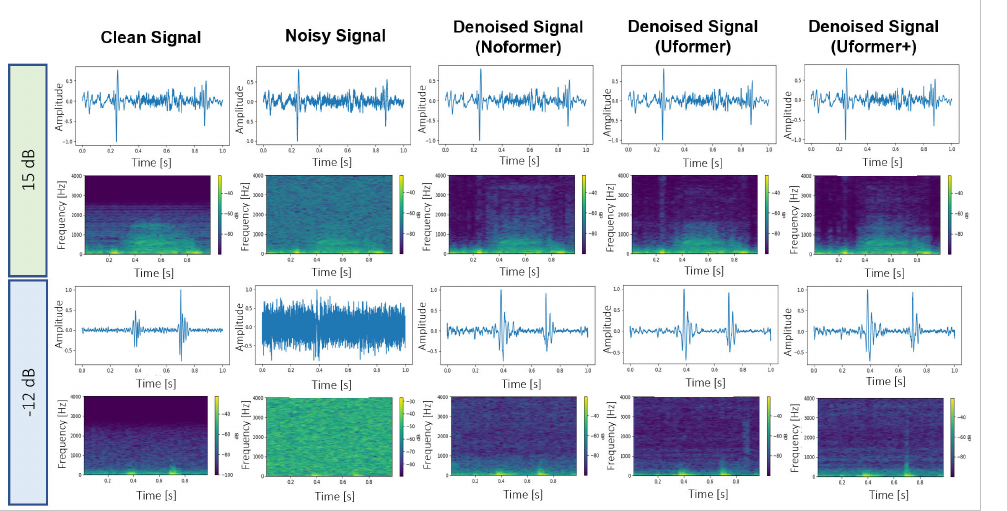}

    \caption{Time–Frequency comparison of clean, noisy, and denoised signals (Noformer, Proposed Uformer, and Uformer+) at two SNR Levels:-12 dB and 15 dB}

    \label{fig:TF}
\end{figure*}

\begin{figure*}
    \centering
    \renewcommand{\arraystretch}{1}
    \includegraphics[width=1\textwidth]{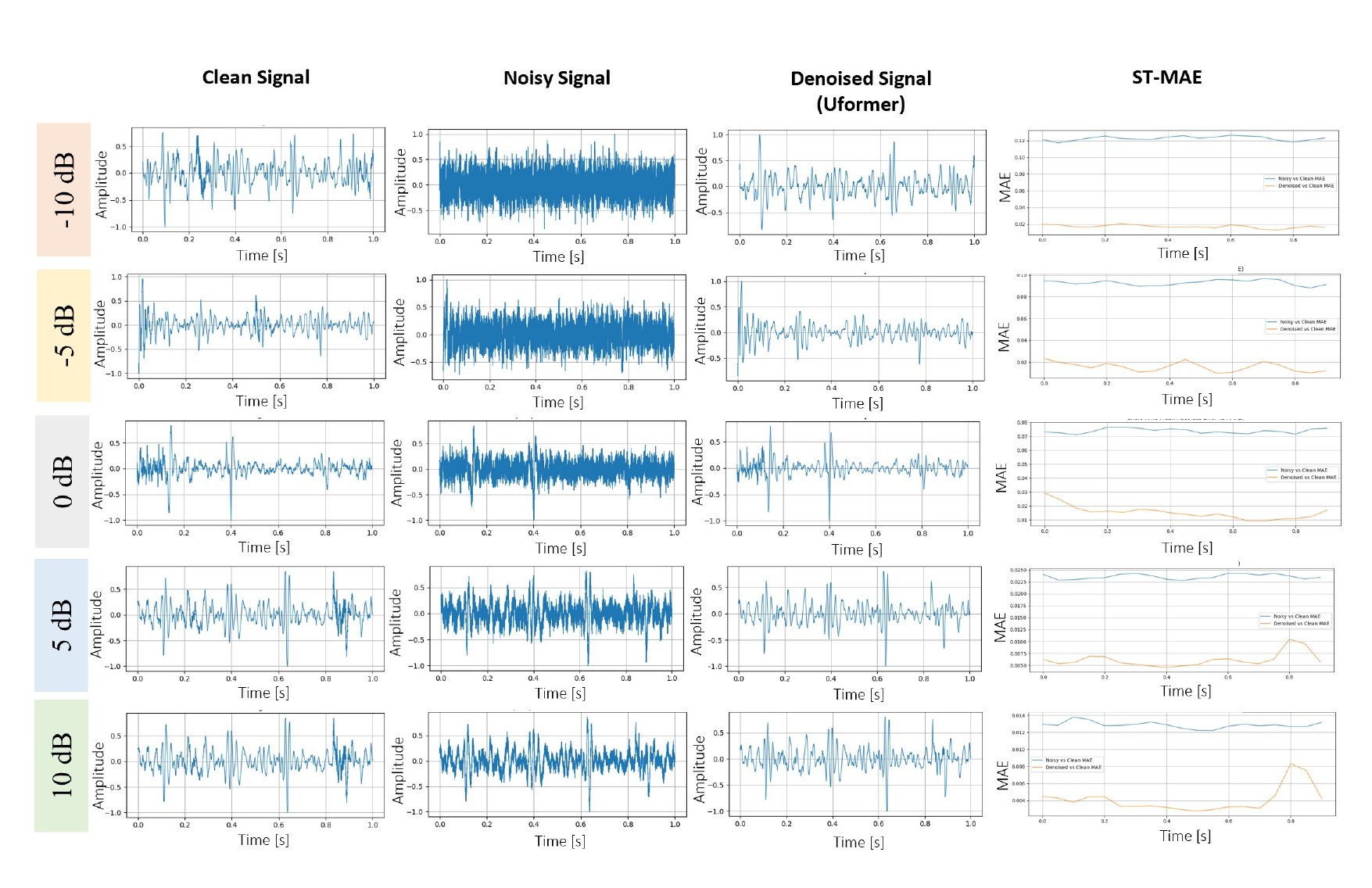}

    \caption{Qualitative and Instantaneous Deviation analysis using waveforms and ST-MAE at multiple input SNR levels for noisy and Uformer's denoised signals}

    \label{fig:ST}
\end{figure*}

\section{Results \& Discussion}\label{sec5}

\subsection{Ablation Study: Effect of the Transformer Encoder}

The central question is whether models with varying attention configurations, specifically the proposed Uformer model, are optimal for this task. To assess the impact of the attention mechanisms, we use the following three model configurations for comparison:

\begin{itemize}
    \item Uformer+: This model incorporates two attention blocks.
    \item Uformer: This version uses a single attention block.
    \item Noformer: As the name suggests, this model operates without any attention blocks.
\end{itemize}
\subsubsection{Quantitative Analysis}
The quantitative performance of these models was assessed across various noise levels, with metrics such as Predicted SNR (PRED\_SNR), PRD, and RMSE used for evaluation using noisy LS data from ICBHI-1, ICBHI\_inf-1 for training and testing purposes, respectively.

The experimental results revealed notable distinctions among the three model configurations at various noise levels. Significantly, the Uformer model, featuring a single attention block, consistently outperformed the other configurations in terms of signal-to-noise ratio (SNR) improvement, as shown in Fig.~\ref{fig:SNR_Im}. This superiority was evident across all SNR levels, highlighting the Uformer's impressive SNR improvement performance. This trend of Uformer's dominance persisted across all noise levels, demonstrating its robustness and effectiveness across all performance criteria, i.e., PRD and RMSE.

\begin{figure}
    \centering
    \renewcommand{\arraystretch}{1}
    \includegraphics[width=1\linewidth]{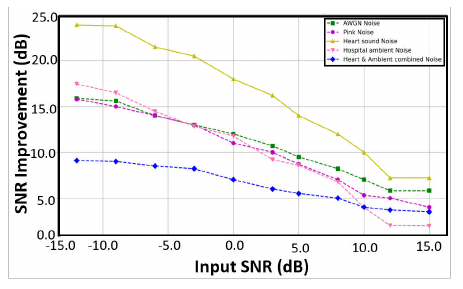}

    \caption{SNR improvement across noise types for different input SNR levels}

    \label{fig:div}
\end{figure}

Overall, the Uformer model exhibited substantial gains, with  4.142\% higher SNR improvement, 0.647\% lower PRD, and 1.698\% lower RMSE on average when compared to Uformer+. Similarly, when compared with Noformer architecture, Uformer outperformed it with  4.882\% higher SNR improvement, 0.967\% lower PRD, and 2.305\% lower RMSE on average (see Table ~\ref{ablation}). These results indicated that the Uformer model, equipped with a single attention block, delivered the best denoising performance among the three configurations. This also suggests that while attention mechanisms can be advantageous for denoising tasks, the optimal number of attention blocks is crucial in achieving superior performance. The double attention block (Uformer+) and the model without any attention (Noformer) were found to be less effective in comparison to the model with a single attention block (Uformer).

\subsubsection{Qualitative Time–Frequency Analysis}
 As shown in Fig. 5, waveform and spectrogram comparisons at extreme SNR conditions (–12 dB and 15 dB) highight the perceptual impact of the attention mechanism. The Noformer configuration leaves noticeable broadband noise, which masks important spectral detail. The Uformer+ variant removes more noise but occasionally over-suppresses subtle spectral features. In contrast, the proposed Uformer achieves the best balance, effectively removing broadband interference while preserving the low- and mid-frequency structures essential for lung sound interpretation. The harmonic ridges and transient bursts in its spectrograms remain sharper and more continuous, indicating superior preservation of clinically relevant information. We also evaluated short-time mean absolute error (ST-MAE) between the clean and Uformer’s denoised signals using a 100 ms frame length with 50\% overlap. As shown in Fig.~\ref{fig:ST}, at low input SNRs (–10 dB, –5 dB), noisy inputs exhibit large error peaks aligned with high-energy respiratory transients. After denoising, Uformer’s ST-MAE curves flatten substantially, indicating reduced instantaneous deviation while preserving waveform morphology. At higher SNRs (5 dB, 10 dB), the error remains near the noise floor for most of the signal duration, reflecting minimal distortion.

Based on the experimental findings, the Uformer model with a single attention block emerges as the most optimal choice to denoise noise-corrupted lung sounds. It also underscores the efficacy of self-attention mechanisms in deep learning and the importance of configuring attention blocks appropriately to achieve optimal results.

\subsection{Performance on Diverse Noise Levels and Noise Sources}

In order to evaluate the effectiveness of the proposed denoising model under various noise conditions, in this section, two primary research questions are answered:  1) Does the model generalize across different noise types? and 2) How robust is the model when confronted with noises originating from different sources? The subsequent subsections are dedicated to providing answers to these questions.

In order to answer the questions, we evaluated the proposed model across different SNRs. Originally, the Uformer model was trained using the ICBHI-1, ICBHI-2, ICBHI-3, and ICBHI-4, which were corrupted with -10dB,-5dB, 0dB, 5dB, 10db, and 15dB. However, the evaluation expands beyond these levels to include unseen noise levels. Moreover, the Uformer model was trained and evaluated using different types of noise, such as synthetic noise (pink noise) and real-world noise, such as: combined heart sound noise and hospital noise ( ICBHI\_inf-2, ICBHI\_inf-3, and ICBHI\_inf-4).

\begin{figure*}
    \centering
    \renewcommand{\arraystretch}{0.85}
    \includegraphics[width=1\textwidth]{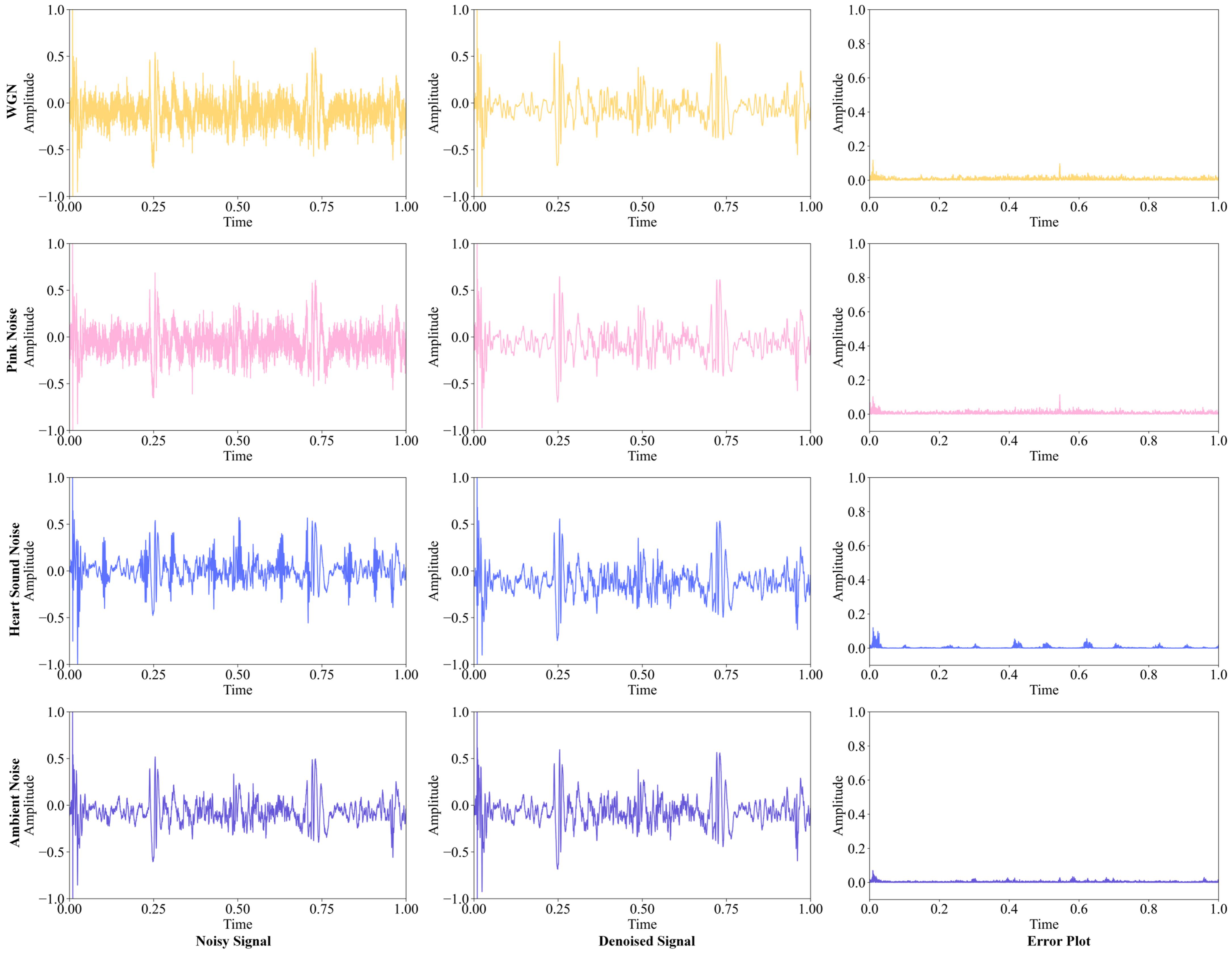}

    \caption{First column contains plots of LS signal (Fig.~\ref{fig:sample}(a)) that have been corrupted with 'White Gaussian Noise (WGN)’, ‘Heart Sound Noise’, 'Pink Noise', and 'Ambient Noise’ respectively. The second column represents the corresponding denoised signals of the first column’s signals using the proposed 'Uformer’ architecture. The third column represents the error plots, which demonstrate the difference between the noisy signals and the denoised signals.}

    \label{fig:sig}
\end{figure*}

Fig \ref{fig:div} demonstrates a clear inverse relationship between input SNR and the achieved SNR improvement across all noise types. Specifically, the denoising performance of the model is most pronounced under low input SNR conditions, indicating its robustness in high-noise environments. Among the evaluated noise types, heart sound noise consistently yields the highest SNR improvement, highlighting the model’s effectiveness in suppressing structured physiological noise. Conversely, performance degrades gradually as the input SNR increases, with diminishing returns at higher SNR levels. Notably, the model achieves moderate and comparable improvements for both AWGN and pink noise, reflecting its adaptability to general stochastic noise types. In contrast, hospital ambient noise initially results in significant SNR gains, but these gains decline rapidly with increasing input SNR, suggesting a limitation in handling non-stationary background noise at cleaner conditions. The most challenging scenario involves heart and ambient noise combined, where the model records the lowest improvements throughout, indicating compounded complexity in multi-source interference suppression. \\
In the results presented, Table~ \ref{RQ3} enumerates the Uformer model's performance over a range of noise levels and types. The model achieved an average SNR improvement of 10.86 dB when assessed with pink noise. In tests involving heart sound and ambient noise, two common real-world noise sources, the Uformer model yielded SNR improvements of 16.41 dB and 10.36 dB, respectively. And for combined noise case it dropped to 6.46 dB This improvement signifies the performance of the model for different types of noises. Fig.~\ref{fig:sig} visually represents various noisy signals, their corresponding denoised outputs generated by the Uformer model, and error plots illustrating the residual differences between the clean and denoised signals. From the figure, it is evident that the Uformer model effectively reduces noise across various types, with the denoised signals closely mirroring the clean signals. The minimal residuals in the error plots further underscore the model's proficiency in handling diverse noise conditions. A synthesis of the visual and numerical data clearly attests to the Uformer model's capability to denoise diverse signal types effectively.
\begin{table*}[!ht]
    \centering
    \caption{Performance of the proposed model at different SNR with various types of noise } \label{RQ3}
    \renewcommand{\arraystretch}{1.4}
    \resizebox{\textwidth}{!}{%
    \begin{tabular}{c|c|c|c|c|c|c|c|c|c}
    \hline\hline
   \multirow{3}{*}{\parbox[c]{2cm}{\centering \textbf{Noise Level \\ (in dB)}}} &\multicolumn{9}{c}{\textbf{Noise Type}}\\
    \cline{2-10}
    
    & \multicolumn{3}{c|}{Pink Noise} & \multicolumn{3}{c|}{Heart sound Noise}  & \multicolumn{3}{c}{Heart + Hospital Noise} \\ 
    
    \cline{2-10}
        & PRED\_SNR & PRD & RMSE & PRED\_SNR & PRD & RMSE & PRED\_SNR & PRD & RMSE \\ 
        \hline \hline
        -12 & 4.582 & 0.971 & 0.058 & 11.625 & 0.953 & 0.027 & 4.624 & 0.969 & 0.062 \\ 
        -10 & 5.697 & 0.954 & 0.051 & 13.726 & 0.938 & 0.021 & 5.395 & 0.952 & 0.056 \\ 
        -8 & 6.855 & 0.930 & 0.045 & 13.893 & 0.915 & 0.020 & 5.971 & 0.928 & 0.052 \\ 
        -5 & 8.671 & 0.872 & 0.036 & 16.255 & 0.863 & 0.016 & 6.773 & 0.870 & 0.048 \\ 
        -2 & 10.520 & 0.783 & 0.029 & 16.687 & 0.779 & 0.015 & 7.561 & 0.783 & 0.043 \\ 
        0 & 11.742 & 0.707 & 0.025 & 18.286 & 0.705 & 0.013 & 8.130 & 0.707 & 0.040 \\ 
        3 & 13.480 & 0.578 & 0.020 & 18.470 & 0.577 & 0.012 & 9.141 & 0.579 & 0.035 \\ 
        5 & 14.568 & 0.490 & 0.017 & 19.900 & 0.490 & 0.011 & 9.920 & 0.490 & 0.032 \\ 
        8 & 16.091 & 0.370 & 0.014 & 19.832 & 0.370 & 0.010 & 11.286 & 0.370 & 0.027 \\ 
        10 & 17.039 & 0.302 & 0.012 & 21.202 & 0.302 & 0.009 & 12.276 & 0.302 & 0.024 \\ 
        12 & 17.936 & 0.244 & 0.011 & 20.727 & 0.244 & 0.009 & 13.322 & 0.244 & 0.020 \\ 
        15 & 19.135 & 0.175 & 0.009 & 22.270 & 0.175 & 0.008 & 14.870 & 0.175 & 0.017 \\ \hline\hline
    \end{tabular}}
\end{table*}
\begin{table}[!ht]
    \centering
    \caption{Performance of the proposed model on independent test sets (AUH) corrupted with WGN noise }\label{unseen}
    \renewcommand{\arraystretch}{1}
\resizebox{\linewidth}{!}{
    \begin{tabular}{c|c|c|c}
    \hline
    \hline
        \parbox[c]{2 cm}{\centering \textbf{Noise Level (in dB)}} & \textbf{PRED\_SNR} & \textbf{PRD} & \textbf{RMSE} \\ \hline\hline
        
        -12 & 6.3525 & 0.9732 & 0.0659 \\ 
        -10 & 7.6549 & 0.9560 & 0.0570 \\ 
        -8 & 8.9132 & 0.9318 & 0.0489 \\ 
        -5 & 10.8448 & 0.8738 & 0.0390 \\ 
        -2 & 12.9760 & 0.7845 & 0.0307 \\ 
        0 & 14.0574 & 0.7080 & 0.0271 \\ 
        3 & 15.8440 & 0.5789 & 0.0219 \\ 
        5 & 16.9562 & 0.4905 & 0.0193 \\ 
        8 & 18.7996 & 0.3704 & 0.0157 \\ 
        10 & 19.8574 & 0.3019 & 0.0140 \\ 
        12 & 20.9797 & 0.2437 & 0.0123 \\ 
        15 & 22.4349 & 0.1754 & 0.0105 \\ \hline\hline
    \end{tabular}}
    
    % }
\end{table}

The results presented in this section indicate that the proposed Uformer model is robust across various SNR levels and different types of noise, making it promising for real-world application. Its performance in mitigating both heart sound and ambient noise further emphasizes its utility in different noisy environments.

\subsection{Performance on Cross-Dataset}

\begin{table*}[!ht]
    \centering
    \caption{Performance of the proposed model on independent test sets (AUH) at different SNR with various types of noise } \label{unseen2}
    \renewcommand{\arraystretch}{1.4}
    \resizebox{\textwidth}{!}{%
    \begin{tabular}{c|c|c|c|c|c|c|c|c|c}
    \hline\hline
   \multirow{3}{*}{\parbox[c]{2cm}{\centering \textbf{Noise Level \\ (in dB)}}} &\multicolumn{9}{c}{\textbf{Noise Type}}\\
    \cline{2-10}
    
    & \multicolumn{3}{c|}{Pink Noise} & \multicolumn{3}{c|}{Heart sound Noise}  & \multicolumn{3}{c}{Heart + Hospital Noise} \\ 
    
    \cline{2-10}
        & PRED\_SNR & PRD & RMSE & PRED\_SNR & PRD & RMSE & PRED\_SNR & PRD & RMSE \\ 
        \hline \hline
-12 & 6.151 & 0.971 & 0.060 & 10.983 & 0.953 & 0.027 & 6.282 & 0.969 & 0.062 \\
-10 & 7.342 & 0.953 & 0.051 & 13.182 & 0.938 & 0.021 & 7.894 & 0.952 & 0.056 \\
 -8 & 8.617 & 0.930 & 0.046 & 14.311 & 0.915 & 0.020 & 9.087 & 0.928 & 0.052 \\
 -5 &10.312 & 0.872 & 0.037 & 16.382 & 0.863 & 0.016 &10.801 & 0.870 & 0.048 \\
 -2 &12.117 & 0.783 & 0.031 & 18.555 & 0.779 & 0.015 &12.457 & 0.783 & 0.043 \\
  0 &13.812 & 0.707 & 0.025 & 20.203 & 0.705 & 0.013 &13.903 & 0.707 & 0.040 \\
  3 &15.273 & 0.578 & 0.020 & 21.203 & 0.577 & 0.011 &15.211 & 0.579 & 0.035 \\
  5 &16.902 & 0.490 & 0.017 & 22.102 & 0.490 & 0.009 &16.873 & 0.490 & 0.032 \\
  8 &18.367 & 0.370 & 0.013 & 23.645 & 0.370 & 0.007 &18.091 & 0.370 & 0.027 \\
 10 &19.406 & 0.302 & 0.011 & 24.806 & 0.302 & 0.006 &19.276 & 0.302 & 0.020 \\
 12 &21.051 & 0.244 & 0.009 & 25.812 & 0.244 & 0.006 &20.129 & 0.244 & 0.017 \\
 15 &22.108 & 0.175 & 0.008 & 26.904 & 0.175 & 0.005 &21.650 & 0.175 & 0.015 
\\ \hline\hline
    \end{tabular}}
\end{table*}

This section evaluates the generalizability of the proposed model on an independent dataset (AUH), which was not seen during training. The model was trained exclusively on the ICBHI dataset and tested on AUH under various realistic noise conditions.

To emulate clinically relevant acoustic disturbances, four types of additive noise were introduced: white Gaussian noise (WGN), pink noise, heart sounds, and combination of heart sound and hospital ambient noise. The model was evaluated at multiple SNR levels ranging from –12 dB to 15 dB. Consistent with expectations, the model exhibited improved PRED\_SNR and decreased PRD and RMSE as the input SNR increased across all noise conditions.

Despite the inherent domain shift, the model maintained comparable, in some cases improved, performance on AUH relative to ICBHI. For example, under WGN contamination, the average PRED\_SNR of AUH dataset was 1.76 dB better compared to ICBHI, confirming domain-invariant denoising efficacy (as shown in Table~\ref{unseen}).

Table~\ref{unseen2} further presents a breakdown of the model’s performance under pink noise, heart sound interference, and combined heart and hospital noise. These results reinforce the model's ability to generalize across both synthetic and physiologically realistic noise types in unseen environments.

Notably, the model achieved its best cross-dataset performance under heart sound noise, despite its spectral overlap with lung sounds. On AUH, it outperformed ICBHI with a higher average PRED\_SNR (20.93 dB vs. 18.56 dB), lower PRD (0.5396 vs. 0.5683), and reduced RMSE (0.0230 vs. 0.0253), indicating superior preservation of lung-specific acoustic structures while suppressing structured interference. Similar trend is also observed for other noise types.

To statistically assess generalizability, a two-sample hypothesis test on PRED\_SNR between AUH and ICBHI across all noise types yielded p-values exceeding 0.8, indicating no significant difference. These findings confirm the model’s robustness and strong generalization across different datasets and noise conditions, underscoring its suitability for real-world deployment in automated lung sound denoising applications.

\subsection{Comparison with State-of-the-art Models}
    \begingroup
\setlength{\tabcolsep}{12pt} % Default value: 6pt
\renewcommand{\arraystretch}{1.8} 
\begin{table*}[h!]
    \centering
\caption{Performance comparison with other SOTA methods}\label{comparison}

    \resizebox{\textwidth}{!}{
    \renewcommand{\arraystretch}{1.2}
        \begin{tabular}{c|c|c|c||c|c|c}
          \hline
          \hline
          \textbf{Noise type} & \multicolumn{3}{c||}{Ambient Noise}&  \multicolumn{3}{c}{Synthetic Noise(WGN)}\\
          \hline
          \hline
          \parbox[c]{1cm}{\centering\textbf{Noise Level}} & \parbox[c]{1.5cm}{\centering \textbf{\textcolor{white}{Dataset} NLMS + Conv-TasNet\cite{yang2023cardiopulmonary} \textcolor{white}{yang2023cardiopulmonary}}} & \parbox[c]{1.5cm}{\centering \textbf{\textcolor{white}{Dataset} NLMS +DINC-Net\cite{yang2023cardiopulmonary}
          \textcolor{white}{yang2023cardiopulmonary}}} & \parbox[c]{2.5cm}{\centering \textbf{Uformer (Proposed Network)}} &  \parbox[c]{1.5cm}{\centering \textbf{IND-M \cite{pouyani2022lung}}}&\parbox[c]{1.5cm}{\centering \textbf{COM-M \cite{pouyani2022lung}}}& \parbox[c]{2.5cm}{\centering \textbf{Uformer (Proposed Network)}}
          \\
                    \hline
          \hline 
          -6 & -2.229 & 7.384 & \textbf{8.467}&  

- &- & -\\
          -3 & -1.223 & 8.536 & \textbf{9.84}&  

- &- & -\\
          0 & 0.09 & 9.618 & \textbf{11.764} &  

9.18
&8.68& \textbf{12.450} \\
          3 & 1.733 & 10.581 & \textbf{12.204} &  

- &- & -\\
          5 & - & - & -&  

12.74
&12.37 & \textbf{15.252} \\
          6 & 3.502 & 11.421 & \textbf{14.571}&  - &- & - \\
          10 & - & - & -&  16.24&15.51& \textbf{19.478} \\
          15 & - & - & -&  

20.2 &17.45& \textbf{21.053} \\
\hline
\hline
\end{tabular}}

\end{table*}
\endgroup

In evaluating the Uformer model's performance and suitability for practical applications, a pivotal aspect is how it fares compared to other state-of-the-art (SOTA)  architectures. This section aims to assess the Uformer model's performance compared to existing methodologies.

We extended our evaluation to include noise levels not initially used in our testing datasets to ensure a fair comparison. Specifically, we incorporated -6 dB, -3 dB, and 6 dB noisy signals, as used by Yang et al. in their work \cite{yang2023cardiopulmonary}. 

% To evaluate the model's efficacy, we relied on the PRED\_SNR matric like other studies.

Table~ \ref{comparison} provides a clear representation of the Uformer model's performance across various SNR levels compared to existing approaches. 
It is important to note that the results cited from \cite{pouyani2022lung} and \cite{yang2023cardiopulmonary} were not obtained using our datasets or evaluation pipeline. These comparisons are included solely as high-level contextual references to illustrate the performance range reported in earlier studies. Due to the unavailability of datasets and insufficient implementation details (e.g., preprocessing steps, network configurations, and evaluation protocols), we were unable to reproduce these methods under consistent experimental conditions for a direct and fair comparison.

The existing methods primarily address two types of noise: ambient noise, as studied by Yang et al. \cite{yang2023cardiopulmonary}, and synthetic noise (WGN), as examined by Pouyani et al. \cite{pouyani2022lung}. The Uformer model outperforms both existing methods across all tested noise levels.

In the study by Pouyani et al. \cite{pouyani2022lung}, they propose two models: individual (IND-M) and combined (COM-M). The IND-M model achieves a 7.09 dB SNR improvement, surpassing the 6.00 dB improvement in COM-M. However, the Uformer model, with a 9.558 dB SNR improvement, outperforms both by 34.80\% and 59.23\%, respectively. Notably, it is essential to point out that IND-M's training and evaluation procedure, where the network is trained with one SNR and evaluated with another, may not be realistic in practical scenarios.

In the study by Yang et al. \cite{yang2023cardiopulmonary}, they experiment with two deep learning models: NLMS + Conv-TasNet and NLMS + DINC-Net, evaluating their performance at various SNR levels (-6 dB, -3 dB, 0 dB, 3 dB, and 6 dB). The NLMS + Conv-TasNet model performs poorly, with an average SNR improvement of 0.375 dB. In contrast, our proposed Uformer model significantly outperforms their best-performing model, NLMS-DINC-Net, achieving an average SNR improvement of 11.369 dB, which is nearly 20\% improvement exhibited by NLMS-DINC. It is worth noting that Yang et al.'s study mainly considers office, hallway, street, and similar noises as ambient noise types, while in reality, lung sounds are corrupted by both physiological noises like heart sounds and ambient noise from the hospital environment. Our study takes this more realistic scenario into account. 

The results indicate that the proposed Uformer model outperforms existing SOTA architectures in denoising lung sounds. Its significant improvements across a range of SNR levels demonstrate its superior performance, making it a promising choice for practical applications in the field.

\subsection{Computational Complexity and Time Performance}

To assess the real-time viability of our proposed model, we conducted a detailed analysis of computational complexity and inference latency across four model variants: Noformer, Uformer, and Uformer+. We evaluated the models using 1-second lung sound segments (8000 samples) and measured three key metrics: total number of trainable parameters, floating point operations (FLOPs), and average inference time per sample. Inference time was measured over 30 repeated runs, excluding the top and bottom 5 outliers for stability.

\begin{table}
\caption{Comparison of computational complexity and inference time across model variants}

\centering
    \renewcommand{\arraystretch}{1.4}
    \resizebox{\linewidth}{!}{%
\begin{tabular}{l|c|c|c}
\hline
\hline
\textbf{Model} & \textbf{Parameters} & \parbox[c]{1.5cm}{\centering\textbf{FLOPs (MFLOPs)}}
 & \parbox[c]{1.5cm}{\centering\textbf{Inference Time (ms/sample)} }

\\
\hline
\hline
Noformer       & 1,131,889           & 2065.55                  & 13.1 \\
Uformer        & 1,268,553           & 2085.37                  & 13.8 \\
Uformer+       & 1,405,217           & 2105.19                  & 17.3 \\
\hline
\hline
\end{tabular}}
\label{tab:complexity_runtime}
\end{table}

As shown in Table~\ref{tab:complexity_runtime}, increasing the number of attention blocks results in a modest increase in both the complexity of the model and the inference time. The total parameters range from 1.13 million (Noformer) to 1.4 million (Uformer+), while the FLOPs increase linearly from 2065.55 to 2125.01 MFLOPs. The inference time remains low across all configurations, ranging from 13.1 to 17.3 ms per sample, demonstrating that all models can easily meet real-time requirements. Among the variants, the proposed Uformer achieves a strong balance between complexity and latency, with a minimal increase in runtime over the baseline (13.8 ms vs. 13.1 ms) while enabling significantly improved denoising performance, as demonstrated in previous sections. This indicates that our model not only offers an end-to-end solution but is also a more suitable choice for integration into edge devices due to its efficiency. Furthermore, when comparing model sizes, our Uformer model has 2.00M (3.27M - 1.27M) fewer parameters than the NLMS-DINC-Net\cite{yang2023cardiopulmonary}. These results confirm that our attention-based denoising framework is computationally efficient and suitable for real-time clinical applications.

\section{Limitations}\label{sec6}

Despite the promising results of the proposed Uformer model in denoising lung sounds under various synthetic and real-world noise conditions, several limitations must be acknowledged. Firstly, a significant portion of the evaluation relies on synthetically added noise at controlled SNR levels, which may not fully capture the variability and complexity of actual clinical environments. Although real-world noise sources such as heart sounds and hospital ambient sounds were considered, the datasets used may not represent the full diversity of acoustic characteristics encountered in different healthcare settings globally. In addition, we observed that the model occasionally fails to preserve subtle pathological components when there is a strong spectral and temporal overlap between lung sounds and complex non-stationary noise, such as human speech or crying infants. In these cases, the denoiser may inadvertently suppress relevant acoustic features, leading to over-smoothing or partial distortion. The performance of the model also tends to degrade in extreme low-SNR conditions, particularly when the pathological sounds are weak and embedded within dominant ambient interference. These scenarios highlight the limitations of attention-driven architectures when applied to severely corrupted signals without additional contextual cues. Although our model was trained on diverse datasets (ICBHI and AUH), we observed a slight drop in performance on out-of-distribution recordings containing unfamiliar noise types or device-specific artifacts, indicating the need for domain adaptation strategies. Furthermore, although the model is designed for real-time processing, its performance has not yet been rigorously validated on resource-constrained platforms such as embedded or wearable hardware. Future studies involving real-time deployments in clinical or telemedicine settings will be necessary to assess the effectiveness of the model to aid clinical decision making and improve diagnostic precision.

\section{Conclusions}\label{sec7}

This study addresses the persistent challenge of noise contamination in lung auscultation signals acquired via electronic stethoscopes. Lung sound recordings are frequently degraded by both convolutional and additive noise sources, which impede accurate clinical interpretation and hinder the early diagnosis of respiratory conditions. Given the importance of high-fidelity signal acquisition, effective denoising is a crucial preprocessing step in lung sound analysis. Traditional denoising methods often introduce artifacts or lead to signal distortion. To overcome these limitations, we proposed the Uformer, a self-attention-based end-to-end neural network for lung sound denoising. The Uformer architecture integrates a two-stage encoder, comprising a convolutional neural network and a transformer encoder, which together facilitate efficient feature extraction and the modeling of long-range dependencies. A CNN-based decoder reconstructs the clean signal, yielding high-quality outputs suitable for clinical interpretation. Our ablation experiments demonstrate that Uformer outperforms its variants, Uformer+ and Noformer, achieving average SNR improvements of 4.142\% and 4.882\%, respectively. Notably, the greatest performance gains were observed under low-SNR conditions, with improvements of up to approximately 16 dB, affirming the model’s robustness in acoustically challenging environments. Comprehensive evaluations were conducted using both synthetic and real-world noise scenarios. Quantitative metrics, including SNR, PRD, and RMSE, were employed to assess overall signal fidelity. Additionally, time–frequency analyses using waveform and spectrogram representations revealed that Uformer effectively attenuates broadband noise while preserving diagnostically important low- and mid-frequency components. This contrasts with the over-suppression observed in Uformer+ and the residual noise retained in Noformer. Further comparisons with conventional band-pass filtering confirmed that the improvements stem from the model’s learned representations rather than simple frequency-based exclusion. The generalization capability of the proposed model was further validated through cross-dataset evaluation. Results showed no statistically significant performance degradation when applied to an unseen clinical dataset, demonstrating strong domain invariance. Compared to existing state-of-the-art denoising approaches, Uformer consistently achieved superior performance across all evaluation metrics. In future, we aim to extend the applicability of Uformer to other physiological auscultation signals, such as bowel sounds. Future research will also explore adaptive attention mechanisms and hardware optimization strategies to enable real-time, low-power deployment in clinical and point-of-care settings. Given its demonstrated effectiveness and versatility, Uformer represents a promising tool for integration into next-generation electronic stethoscope systems and automated diagnostic platforms.

\bibliographystyle{unsrtnat}   % or plain, ieeetr, etc.
\bibliography{Uformer}

\begin{thebibliography}{42}
\providecommand{\natexlab}[1]{#1}
\providecommand{\url}[1]{\texttt{#1}}
\expandafter\ifx\csname urlstyle\endcsname\relax
  \providecommand{\doi}[1]{doi: #1}\else
  \providecommand{\doi}{doi: \begingroup \urlstyle{rm}\Url}\fi

\bibitem[Newsroom(2023)]{ihme2023}
IHME Newsroom.
\newblock Chronic respiratory disease is third leading cause of death globally with air pollution killing 1.3 million people.
\newblock IHME Newsroom, 4 2023.
\newblock URL \url{https://www.healthdata.org/news-events/newsroom/news-releases/chronic-respiratory-disease-third-leading-cause-death-globally}.
\newblock Accessed: 2023-12-14.

\bibitem[Reichert et~al.(2008)Reichert, Gass, Brandt, and Andr{\`e}s]{reichert2008analysis}
Sandra Reichert, Raymond Gass, Christian Brandt, and Emmanuel Andr{\`e}s.
\newblock Analysis of respiratory sounds: state of the art.
\newblock \emph{Clin. Med. Circ. Respirat. Pulm. Med.}, 2:\penalty0 CCRPM--S530, 2008.

\bibitem[Lozano et~al.(2016)Lozano, Fiz, and Jan{\'e}]{lozano2016performance}
Manuel Lozano, Jos{\'e}~Antonio Fiz, and Raimon Jan{\'e}.
\newblock Performance evaluation of the hilbert--huang transform for respiratory sound analysis and its application to continuous adventitious sound characterization.
\newblock \emph{Signal Process.}, 120:\penalty0 99--116, 2016.

\bibitem[Bohadana et~al.(2014)Bohadana, Izbicki, and Kraman]{bohadana2014fundamentals}
Abraham Bohadana, Gabriel Izbicki, and Steve~S Kraman.
\newblock Fundamentals of lung auscultation.
\newblock \emph{N. Engl. J. Med.}, 370\penalty0 (8):\penalty0 744--751, 2014.

\bibitem[Gurung et~al.(2011)Gurung, Scrafford, Tielsch, Levine, and Checkley]{gurung2011computerized}
Arati Gurung, Carolyn~G Scrafford, James~M Tielsch, Orin~S Levine, and William Checkley.
\newblock Computerized lung sound analysis as diagnostic aid for the detection of abnormal lung sounds: a systematic review and meta-analysis.
\newblock \emph{Respir. Med.}, 105\penalty0 (9):\penalty0 1396--1403, 2011.

\bibitem[Bhosale and Patnaik(2023)]{bhosale2023puldi}
Yogesh~H Bhosale and K~Sridhar Patnaik.
\newblock Puldi-covid: Chronic obstructive pulmonary (lung) diseases with covid-19 classification using ensemble deep convolutional neural network from chest x-ray images to minimize severity and mortality rates.
\newblock \emph{Biomed. Signal Process. Control}, 81:\penalty0 104445, 2023.

\bibitem[Shuvo and Mamun(2023)]{shuvo2023automated}
Samiul~Based Shuvo and Tasnia~Binte Mamun.
\newblock An automated end-to-end deep learning-based framework for lung cancer diagnosis by detecting and classifying the lung nodules, April 2023.

\bibitem[Pouyani et~al.(2022)Pouyani, Vali, and Ghasemi]{pouyani2022lung}
Mozhde~Firoozi Pouyani, Mansour Vali, and Mohammad~Amin Ghasemi.
\newblock Lung sound signal denoising using discrete wavelet transform and artificial neural network.
\newblock \emph{Biomed. Signal Process. Control}, 72:\penalty0 103329, 2022.

\bibitem[Yang et~al.(2023)Yang, Dai, Wang, Cai, Wang, and Hu]{yang2023cardiopulmonary}
Chunjian Yang, Neng Dai, Zhi Wang, Shengsheng Cai, Jiajun Wang, and Nan Hu.
\newblock Cardiopulmonary auscultation enhancement with a two-stage noise cancellation approach.
\newblock \emph{Biomed. Signal Process. Control}, 79:\penalty0 104175, 2023.

\bibitem[Baharanchi et~al.(2022)Baharanchi, Vali, and Modaresi]{baharanchi2022noise}
Shahrzad~Abbasi Baharanchi, Mansour Vali, and Mohammadreza Modaresi.
\newblock Noise reduction of lung sounds based on singular spectrum analysis combined with discrete cosine transform.
\newblock \emph{Appl. Acoust.}, 199:\penalty0 109005, 2022.

\bibitem[Singh et~al.(2020)Singh, Singh, and Behera]{singh2020comparative}
Divya Singh, Bikesh~Kumar Singh, and Ajoy~Kumar Behera.
\newblock Comparative analysis of lung sound denoising technique.
\newblock In \emph{Proc. ICPC2T}, pages 406--410. IEEE, 2020.

\bibitem[Meng et~al.(2019)Meng, Wang, Shi, and Zhao]{meng2019kind}
Fei Meng, Yixuan Wang, Yan Shi, and Hongmei Zhao.
\newblock A kind of integrated serial algorithms for noise reduction and characteristics expanding in respiratory sound.
\newblock \emph{Int. J. Biol. Sci.}, 15\penalty0 (9):\penalty0 1921, 2019.

\bibitem[Syahputra et~al.(2017)Syahputra, Situmeang, Rahmat, and Budiarto]{syahputra2017noise}
MF~Syahputra, SIG Situmeang, RF~Rahmat, and R~Budiarto.
\newblock Noise reduction in breath sound files using wavelet transform based filter.
\newblock In \emph{IOP MSE}, volume 190, page 012040. IOP Publishing, 2017.

\bibitem[Ding and Selesnick(2015)]{ding2015artifact}
Yin Ding and Ivan~W Selesnick.
\newblock Artifact-free wavelet denoising: Non-convex sparse regularization, convex optimization.
\newblock \emph{IEEE Signal Process. Lett.}, 22\penalty0 (9):\penalty0 1364--1368, 2015.

\bibitem[Ghael et~al.(1997)Ghael, Sayeed, and Baraniuk]{ghael1997improved}
Sadeep Ghael, Akbar~M Sayeed, and Richard~G Baraniuk.
\newblock Improved wavelet denoising via empirical wiener filtering.
\newblock In \emph{SPIE Wavelet Appl. Signal Process.}, 1997.

\bibitem[Gallaire and Sayeed(1998)]{gallaire1998wavelet}
J-PG Gallaire and Akbar~M Sayeed.
\newblock Wavelet-based empirical wiener filtering.
\newblock In \emph{Proc. TFTS}, pages 641--644. IEEE, 1998.

\bibitem[Choi and Baraniuk(1998)]{choi1998analysis}
Hyeokho Choi and Richard Baraniuk.
\newblock Analysis of wavelet-domain wiener filters.
\newblock In \emph{Proc. TFTS}, pages 613--616. IEEE, 1998.

\bibitem[Nikolaev et~al.(2000)Nikolaev, Nikolov, Gotchev, and Egiazarian]{nikolaev2000wavelet}
Nikolay Nikolaev, Z~Nikolov, A~Gotchev, and K~Egiazarian.
\newblock Wavelet domain wiener filtering for ecg denoising using improved signal estimate.
\newblock In \emph{Proc. ICASSP}, volume~6, pages 3578--3581. IEEE, 2000.

\bibitem[Mondal et~al.(2011)Mondal, Bhattacharya, and Saha]{mondal2011reduction}
Ashok Mondal, PS~Bhattacharya, and Goutam Saha.
\newblock Reduction of heart sound interference from lung sound signals using empirical mode decomposition technique.
\newblock \emph{J. Med. Eng. Technol.}, 35\penalty0 (6-7):\penalty0 344--353, 2011.

\bibitem[Molaie et~al.(2014)Molaie, Jafari, Moradi, Sprott, and Golpayegani]{molaie2014chaotic}
Malihe Molaie, Sajad Jafari, Mohammad~Hasan Moradi, Julien~Clinton Sprott, and S~Mohammad Reza~Hashemi Golpayegani.
\newblock A chaotic viewpoint on noise reduction from respiratory sounds.
\newblock \emph{Biomed. Signal Process. Control}, 10:\penalty0 245--249, 2014.

\bibitem[Haider(2021)]{haider2021respiratory}
Nishi~Shahnaj Haider.
\newblock Respiratory sound denoising using empirical mode decomposition, hurst analysis and spectral subtraction.
\newblock \emph{Biomed. Signal Process. Control}, 64:\penalty0 102313, 2021.

\bibitem[Haider et~al.(2018)Haider, Periyasamy, Joshi, and Singh]{haider2018savitzky}
Nishi~Shahnaj Haider, R~Periyasamy, Deepak Joshi, and BK~Singh.
\newblock Savitzky-golay filter for denoising lung sound.
\newblock \emph{Braz. Arch. Biol. Technol.}, 61, 2018.

\bibitem[Emmanouilidou et~al.(2015)Emmanouilidou, McCollum, Park, and Elhilali]{emmanouilidou2015adaptive}
Dimitra Emmanouilidou, Eric~D McCollum, Daniel~E Park, and Mounya Elhilali.
\newblock Adaptive noise suppression of pediatric lung auscultations with real applications to noisy clinical settings in developing countries.
\newblock \emph{IEEE Trans. Biomed. Eng.}, 62\penalty0 (9):\penalty0 2279--2288, 2015.

\bibitem[Al-Naggar and Al-Udyni(2018{\natexlab{a}})]{al2018performance}
Noman~Q Al-Naggar and Mohammed~H Al-Udyni.
\newblock Performance of adaptive noise cancellation with normalized last-mean-square based on the signal-to-noise ratio of lung and heart sound separation.
\newblock \emph{J. Healthc. Eng.}, 2018, 2018{\natexlab{a}}.

\bibitem[Yin et~al.(2023)Yin, Liu, Li, Qian, and Chen]{yin2023gan}
Jin Yin, Aiping Liu, Chang Li, Ruobing Qian, and Xun Chen.
\newblock A gan guided parallel cnn and transformer network for eeg denoising.
\newblock \emph{IEEE J. Biomed. Health Inform.}, 2023.

\bibitem[An et~al.(2022)An, Lam, and Ling]{an2022auto}
Yang An, Hak~Keung Lam, and Sai~Ho Ling.
\newblock Auto-denoising for eeg signals using generative adversarial network.
\newblock \emph{Sensors}, 22\penalty0 (5):\penalty0 1750, 2022.

\bibitem[Singh and Sharma(2022)]{singh2022attention}
Prateek Singh and Ambalika Sharma.
\newblock Attention-based convolutional denoising autoencoder for two-lead ecg denoising and arrhythmia classification.
\newblock \emph{IEEE Trans. Instrum. Meas}, 71:\penalty0 1--10, 2022.

\bibitem[Wang et~al.(2022)Wang, Chen, Zeng, Wang, Liu, Liu, Tian, and Lu]{wang2022ecg}
Xiaoyu Wang, Bingchu Chen, Ming Zeng, Yuli Wang, Hui Liu, Ruixia Liu, Lan Tian, and Xiaoshan Lu.
\newblock An ecg signal denoising method using conditional generative adversarial net.
\newblock \emph{IEEE J. Biomed. Health Inform.}, 26\penalty0 (7):\penalty0 2929--2940, 2022.

\bibitem[Ali et~al.(2023)Ali, Shuvo, Al-Manzo, Hasan, and Hasan]{ali2023end}
Shams~Nafisa Ali, Samiul~Based Shuvo, Md~Ishtiaque~Sayeed Al-Manzo, Anwarul Hasan, and Taufiq Hasan.
\newblock An end-to-end deep learning framework for real-time denoising of heart sounds for cardiac disease detection in unseen noise.
\newblock \emph{IEEE Access}, 2023.

\bibitem[Chandel et~al.(2022)Chandel, Kumar, and Muduli]{chandel2022stacked}
Abhinav Chandel, Vinit Kumar, and Priya~Ranjan Muduli.
\newblock Stacked bi-lstm network and dual signal transformation for heart sound denoising.
\newblock In \emph{Lect. Notes Electr. Eng.}, pages 123--133. Springer, 2022.

\bibitem[Rocha et~al.(2018)Rocha, Filos, Mendes, Vogiatzis, Perantoni, Kaimakamis, Natsiavas, Oliveira, J{\'a}come, Marques, et~al.]{rocha2018alpha}
BM~Rocha, Dimitris Filos, Lea Mendes, Ioannis Vogiatzis, Eleni Perantoni, Evangelos Kaimakamis, P~Natsiavas, Ana Oliveira, C~J{\'a}come, A~Marques, et~al.
\newblock A respiratory sound database for the development of automated classification.
\newblock In \emph{ICBHI 2017}, pages 33--37. Springer, 2018.

\bibitem[Fraiwan et~al.(2021)Fraiwan, Fraiwan, Khassawneh, and Ibnian]{fraiwan2021dataset}
Mohammad Fraiwan, Luay Fraiwan, Basheer Khassawneh, and Ali Ibnian.
\newblock A dataset of lung sounds recorded from the chest wall using an electronic stethoscope.
\newblock \emph{Data Br.}, 35:\penalty0 106913, 2021.

\bibitem[Yaseen et~al.(2018)Yaseen, Son, and Kwon]{yaseen2018classification}
Yaseen, Gui-Young Son, and Soonil Kwon.
\newblock Classification of heart sound signal using multiple features.
\newblock \emph{Applied Sciences}, 8\penalty0 (12):\penalty0 2344, 2018.

\bibitem[Shuvo et~al.(2023)Shuvo, Alam, Ayman, Chakma, Barua, and Acharya]{shuvo2023nrc}
Samiul~Based Shuvo, Syed~Samiul Alam, Syeda~Umme Ayman, Arbil Chakma, Prabal~Datta Barua, and U~Rajendra Acharya.
\newblock Nrc-net: Automated noise robust cardio net for detecting valvular cardiac diseases using optimum transformation method with heart sound signals.
\newblock \emph{Biomedical Signal Processing and Control}, 86:\penalty0 105272, 2023.

\bibitem[Alam et~al.(2023)Alam, Chakma, Rahman, Bin~Mofidul, Alam, Utama, and Jang]{alam2023rf}
Syed~Samiul Alam, Arbil Chakma, Md~Habibur Rahman, Raihan Bin~Mofidul, Md~Morshed Alam, Ida Bagus Krishna~Yoga Utama, and Yeong~Min Jang.
\newblock Rf-enabled deep-learning-assisted drone detection and identification: An end-to-end approach.
\newblock \emph{Sensors}, 23\penalty0 (9):\penalty0 4202, 2023.

\bibitem[El~Helou and Susstrunk(2020)]{el2020blind}
Majed El~Helou and Sabine Susstrunk.
\newblock Blind universal bayesian image denoising with gaussian noise level learning.
\newblock \emph{IEEE Transactions on Image Processing}, 29:\penalty0 4885--4897, 2020.

\bibitem[Zhou et~al.(2020)Zhou, Jiao, Huang, Wang, Wang, Shi, and Huang]{zhou2020awgn}
Yuqian Zhou, Jianbo Jiao, Haibin Huang, Yang Wang, Jue Wang, Honghui Shi, and Thomas Huang.
\newblock When awgn-based denoiser meets real noises.
\newblock In \emph{Proc. AAAI}, volume~34, pages 13074--13081, 2020.

\bibitem[Ren et~al.(2018)Ren, Jin, Chen, Ghayvat, and Chen]{ren2018novel}
Haoran Ren, Hailong Jin, Chen Chen, Hemant Ghayvat, and Wei Chen.
\newblock A novel cardiac auscultation monitoring system based on wireless sensing for healthcare.
\newblock \emph{IEEE J. Transl. Eng. Health Med.}, 6:\penalty0 1--12, 2018.

\bibitem[Nersisson and Noel(2017)]{nersisson2017heart}
Ruban Nersisson and Mathew~M Noel.
\newblock Heart sound and lung sound separation algorithms: a review.
\newblock \emph{J. Med. Eng. Technol.}, 41\penalty0 (1):\penalty0 13--21, 2017.

\bibitem[Liu et~al.(2016)Liu, Springer, Li, Moody, Juan, Chorro, Castells, Roig, Silva, Johnson, et~al.]{liu2016open}
Chengyu Liu, David Springer, Qiao Li, Benjamin Moody, Ricardo~Abad Juan, Francisco~J Chorro, Francisco Castells, Jos{\'e}~Millet Roig, Ikaro Silva, Alistair~EW Johnson, et~al.
\newblock An open access database for the evaluation of heart sound algorithms.
\newblock \emph{Phys. Meas.}, 37\penalty0 (12):\penalty0 2181, 2016.

\bibitem[Kingma and Ba(2014)]{kingma2014adam}
Diederik~P Kingma and Jimmy Ba.
\newblock Adam: A method for stochastic optimization.
\newblock \emph{arXiv preprint arXiv:1412.6980}, 2014.

\bibitem[Al-Naggar and Al-Udyni(2018{\natexlab{b}})]{wilson2017marginal}
Noman~Q Al-Naggar and Mohammed~H Al-Udyni.
\newblock Performance of adaptive noise cancellation with normalized last-mean-square based on the signal-to-noise ratio of lung and heart sound separation.
\newblock \emph{J. Healthc. Eng.}, 2018, 2018{\natexlab{b}}.

\end{thebibliography}
\end{document}